\documentclass[12pt]{article}
\newcommand{\beq}{\begin{equation}}
\newcommand{\eeq}{\end{equation}}
\newcommand{\bdis}{\begin{displaymath}}
\newcommand{\edis}{\end{displaymath}}
\newcommand{\beqa}{\begin{eqnarray}}
\newcommand{\beas}{\begin{eqnarray*}}
\newcommand{\eeqa}{\end{eqnarray}}
\newcommand{\eeas}{\end{eqnarray*}}
\newcommand{\ba}{\begin{array}}
\newcommand{\ea}{\end{array}}
\newcommand{\esp}{\vspace{.1cm}}
\newcommand{\esk}{\vspace{.5cm}}
\newcommand{\hsp}{\hspace{1cm}}
\newcommand{\hlf}{\hspace{.5cm}}
\newcommand{\smn}{\hspace{-1cm}}
\newcommand{\smp}{\hspace{+2.5cm}}

\begin{document}
\input{psfig}
\begin{center}
\begin{Large}
\begin{bf}
A Generalized Uhlenbeck and Beth Formula \\
for the Third Cluster Coefficient \\
\end{bf}
\end{Large}
\vspace{.7cm}
Sigurd Yves Larsen \footnote{Professor Emeritus} \\
 Department of Physics, Temple University,
 Philadelphia, PA 19122, USA, \\
 Monique Lassaut \\
Institut de Physique Nucl\'{e}aire, CNRS-IN2P3, Universit\'e Paris-Sud,\\
Universit\'e Paris-Saclay, F-91406 Orsay Cedex, France \\
 and \\
 Alejandro Amaya-Tapia\\
 Instituto de Ciencias F\'{i }sicas, Universidad Nacional Aut\'onoma de M\'exico,\\
 AP 48-3, Cuernavaca, Mor. 62251, M\'{e}xico.\\

\vspace{1.cm}
\begin{bf}
ABSTRACT \\
\end{bf}
\vspace{.3cm}
\end{center}
Relatively recently (A. Amaya-Tapia, S.  Y. Larsen, M. Lassaut. 
\textit{Ann. Phys.},\textbf{306}
(2011) 406), we presented a formula for the evaluation of 
the third Bose fugacity coefficient - leading to the third virial coefficient -
in terms of three-body eigenphase shifts,
for particles subject to repulsive forces. {\bf An analytical}  calculation for a 1-dim. model,
for which the result is known, confirmed the validity of this approach.
We now extend the formalism to particles with attractive forces, and  
therefore must allow for the possibility that the particles have bound states.
We thus obtain a true generalization of the famous formula 
of Uhlenbeck and Beth (G.E. Uhlenbeck, E. Beth. Physica,\textbf{3} (1936) 729;
E. Beth, G.E. Uhlenbeck. \textbf{ibid}, \textbf{4} (1937) 915) (and of Gropper
(L. Gropper. Phys. Rev. {\bf 50} (1936) 963; \textbf{ibid} \textbf{51} (1937) 
1108)) for the second virial. 
We illustrate our formalism by a calculation, in an adiabatic approximation,
of the third cluster in one dimension, using McGuire's model as in our 
previous paper, but with attractive forces.
The inclusion of three-body bound states is trivial; taking into account 
states having asymptotically two particles bound, and one free, is not.

\newpage
\pagestyle{plain}
\pagenumbering{arabic}
\setcounter{page}{2}
\baselineskip=14pt
\vspace{-2cm}
\begin{center}
\section*{Introduction}
\end{center}
\vspace{.3cm}

Our goal, over many decades, has been to develop a generalization of 
the formula of Uhlenbeck and Beth\cite{uhb}(and also of Gropper\cite{gro}),
which yields the second virial in terms of phase shifts and bound state energies.
This would be an expression for the higher virials, in terms of quantities
which characterize the asymptotic - long range - behaviour of the wave 
functions which appear in an eigenfunction evaluation of the Statistical 
Mechanical traces. 
These would be eigenphase shifts and bound state energies.

\esp 
This effort has led to 
an approach for the third virial using 
an expansion of the wave functions in terms of hyperspherical 
harmonics\cite{lam} -
only made really useful by an adiabatic approximation.
Among other results, one has been able to show that in the semi-classical
limit,  for repulsive forces, one recovers the known classical results
from an eigenphase shift formulation\cite{lpb}.
See, also, the calculation of the third virial in two dimensions, 
for repulsive step potentials\cite{laz,kil,tony},
and - as a bonus but important - a formulation of the second virial 
for particles subject to anisotropic forces\cite{lap}, i.e., for 
a Helium atom and a Hydrogen molecule.

\esp 
This latter formulation, in fact, has been crucial for us. 
It showed how, 
instead of putting particles in a box and calculating a density of 
states, one could use a procedure similar to that used in obtaining 
a Wronskian, and obtain analytically the result of integrating the 
square of the wave functions, in terms of these wave functions 
(and their derivatives) at the origin (giving zero) and at the 
long range limit of integration.
I.e., it permitted us to work in the continuum and to express
our results in terms of asymptotic quantities, precisely eigenphase 
shifts and bound state eigenvalues.

\esp
The next crucial step, for our three particles, towards including the bound 
states - such as having the possibility at long range of a 2 body bound 
state and a free particle - was to take advantage of a 
hyperspherical adiabatic basis\cite{lar}. 

\esp
Its importance is two-fold. 

\esp 
The first is due to the fact that, even for short ranged 2-body potentials, 
the effective few-body interactions are long ranged and so, also, are their 
effects on the continuum wave functions.
The use of the hyperspherical adiabatic basis, which incorporates 
information 
valid for large values of the hyper radius, is tremendously helpful. 

\esp 
The second is that at large distances the free and bound structures are
made explicitly clear\cite{glp} and, in an expansion, are associated with 
distinct amplitudes. 

\esp 
We note, for example, that in a three-body
problem with a 2-body bound state, at least one of the elements of
the adiabatic basis will have, built in,
the cluster property of the 2 bodies, and others will correspond 
to asymptotically free particles.

\esp 
We note also that in both  hyperspherical approaches the hard step
conceptually is to pass from two to three particles. The extension to 
a higher number of particles is straightforward. We limit ourselves 
here to a maximum of three particles, sufficient for the third virial.

\newpage
In a hyperspherical adiabatic reformulation of our formalism, we were
then able, in the absence of bound states, to give a formula expressing the 
quantum mechanical third cluster, in terms of adiabatic phase shifts\cite{lar}. 
(This for Boltzmann statistics.) 
From this q. m. formalism, under the same constraints, we
were again able to recover, as $\hbar$ goes to $0$, the classical 
expression. We also discussed some of the difficulties that arise in the
presence of 2-body bound states and presented a tentative formula involving
eigenphase shifts and 2 and 3 body bound state energies.

\esp
Recently, we generalized our formalism to accommodate Bose statistics
and calcu\-lated\cite{asm} the third fugacity coefficient for a 
version of a model due
to McGuire\cite{mcg}. It consists of 3 particles on a line,
subject to repulsive delta function potentials. 
The model allowed us to obtain many results in analytical form\cite{syj,vlp} 
and to
compare our results with those of Dodd and Gibbs\cite{dog}, who integrated
expressions requiring the complete (known) wave functions for this model. 
This confirmed the validity of our (more general) approach.

\esp
In the present work, we return to the problem of the bound states.
It simplifies our discussion to use particles subject
to quantum statistics (say Bose) and though our formalism is stated 
for three dimensions, we will illustrate the behaviour of certain 
potentials and bound state situations by borrowing results 
from our previous work in one dimension. In addition we present new
results for McGuire's model, this time yielding the virial for
attractive potentials.

\esp
We will see that in the complete generalization, we still encounter certain difficulties and constraints
and we discuss subtleties and fine points in the whole attempt to establish
phase shifts plus bound state formalisms - even for the second virial.

\newpage
\topmargin=0in
\begin{center}
\section*{Statistical Mechanics}
\end{center}

We start with the Grand Partition function:

\begin{equation}
{\cal Q}={\displaystyle \sum_{n=0}^{\infty}z^{n}{}Tr\left(e^{-\beta H_{n}}\right)}=1+zTr\left(e^{-\beta T_{1}}\right)+z^{2}{}Tr\left(e^{-\beta H_{2}}\right)+
z^{3}{}Tr\left(e^{-\beta H_{3}}\right)+\cdots
\label{part}
\end{equation}
 The fugacity $z$ equals $\exp\left(\mu/\kappa T\right)$;$\;\beta=1/\kappa T$,
where $\mu$, $\kappa$ and $T$ are the Gibbs' function per particle,
Boltzmann's constant and the temperature, respectively. $H_{n}$ and
$T_{n}$ are the n-particle Hamiltonian and kinetic energy operators.

We note that no factor of  ($1/n!$) appears in this development.
This is correct for Bose or Fermi statistics. 
Also important, is the fact that \[
Tr^{}\left(e^{-\beta H_{n}}\right)\rightarrow V^{n}\;\; {\rm as}\; 
{\rm the} \; {\rm volume} \; \; V \; \rightarrow \; {\rm large},\]
 leading to the divergence of the individual traces in the thermodynamic
limit. If we take, however, the logarithm of the Grand Partition function
${\cal Q}$, we obtain 
\begin{equation}
\begin{array}{cl}
ln{\cal Q} & =z\,\, Tr\left(e^{-\beta T_{1}}\right)\\
 & +z^{2}\left[Tr\left(e^{-\beta H_{2}}\right)-\frac{1}{2}\left[Tr\left(e^{-\beta T_{1}}\right)\right]^{2}\right]\\
 & +z^{3}\left[Tr\left(e^{-\beta H_{3}}\right)-Tr\left(e^{-\beta T_{1}}\right)Tr\left(e^{-\beta H_{2}}\right)+\frac{1}{3}\left[Tr\left(e^{-\beta T_{1}}\right)\right]^{3}\right]\\
 & +\cdots
\end{array}
\label{lnq}
\end{equation}
 which, when divided by V, gives coefficients of the powers of $z$,
which are independent of the volume, when the latter becomes large.
They are the fugacity coefficients $b_{l}$. 
\esp 
We can then write for the pressure and the density 
\[ p/\kappa T=(1/V)\ln{\cal Q}=\sum_{l}b_{l}\, z^{l},\] \[
N/V=\rho=\sum_{l}l\, b_{l}\, z^{l},\]
 and the fugacity can be eliminated to yield the pressure in terms
of the density, in the virial expansion: 
\[ \frac{p}{\mathrm{k}T}=\sum_{n=1}^{\infty}B_{n}\rho^{n}.\]

\newpage
\subsection*{The third fugacity coefficient, Bose statistics}

\esp
\esp
From Eq.(\ref{lnq}),
an expression for the 3rd fugacity coefficient can be written as 
\begin{equation}
b_{3}=\frac{1}{V}\left[Tr\left(e^{-\beta H_{3}}\right)-
Tr\left(e^{-\beta T_{1}}\right)Tr\left(e^{-\beta H_{2}}\right)+
\frac{1}{3}Tr\left(e^{-\beta T_{1}}\right)^{3}\right]
\label{b3}
\end{equation}
 Note that each of the 3 terms in the bracket, above, grows as $V^{3}$
asymptotically, for large $V$. Therefore, subtractions must take
place, so that $b_{3}$ might become independent of $V$, as is required.
If we factor out the contribution of the C.M. (proportional to $V$),
each term will have a dependence proportional to $V^{2}$.

\esp
To help us decrease the $V$ dependence, let us subtract, from each
of terms above, the equivalent term without interaction.
\esp
Subtracting, therefore, $b_{3}^{0}$ from the $b_{3}$, we obtain:
\begin{equation}
b_{3}-b_{3}^{0}=\frac{1}{V}\left[Tr\left(e^{-\beta H_{3}}-
e^{-\beta T_{3}}\right)-
Tr\left(e^{-\beta T_{1}}\right)Tr\left(e^{-\beta H_{2}}-
e^{-\beta T_{2}}\right)\right] .
\label{b30a}
\end{equation}
\esp
For Boltzmann statistics $b^0_{3}$ equals zero.
For Fermi and Bose statistics $b^0_{3} = 1/(3^{5/2}\lambda^3_T)$,
where $\lambda^3_T = (2\pi\hbar^2 \beta/m)^{3/2}$, and $\lambda_T$ is the 
thermal wave length.
We have ignored the factors associated with spin.

\esp
To subtract volumes from comparable volumes, we rewrite the traces
so that the arguments of the exponentials are all 3-body Hamiltonians:
\esp
\begin{equation}
b_{3}-b_{3}^{0}=\frac{1}{V}\left[Tr\left(e^{-\beta H_{3}}-
e^{-\beta T_{3}}\right)-
Tr\left(e^{-\beta (H_{2} + T_{1})}  -
e^{-\beta (T_{2} + T_{1})}\right)\right] .
\label{b30b}
\end{equation}

\esp
One remark: We will evaluate the traces by inserting complete sets
of states into the traces. For the terms involving $H_{3}$ and $T_{3}$,
we will choose states having complete symmetry between the three particles.
For the terms involving $H_{2}$ or $T_{2}$, above, we will need symmetry 
between say particles 1 and 2, with particle 3 acting as a spectator.
The adiabatic functions that we define 
in the next section will have to satisfy these requirements. Different 
sets of indices will also be associated with each type of symmetry.
 
\newpage{} 
\begin{center}
\section*{Hyperspherical Adiabatic Preliminaries}
\end{center}
\vspace{.3cm}

  For the 3 particles of equal masses, in three dimensions, we first
introduce center of mass and Jacobi coordinates. We define
\beq
 \vec{\eta}  =  \left(\frac{1}{2}\right)^{1/2}(\vec{r}_{1} - \vec{r}_{2})
\hsp
 \vec{\xi}  =   \left(\frac{2}{3}\right)^{1/2}  \left(\frac{\vec{r}_{1} + \vec{r}_{2}}{2}
- \vec{r}_{3}\right)
\hsp \vec{R}  =  \, \; \frac{1}{3} \; (\vec{r}_{1} + \vec{r}_{2} +
\vec{r}_{3})
\eeq
\vspace{.10in}
\begin{figure}[h]
\centerline{\psfig{file=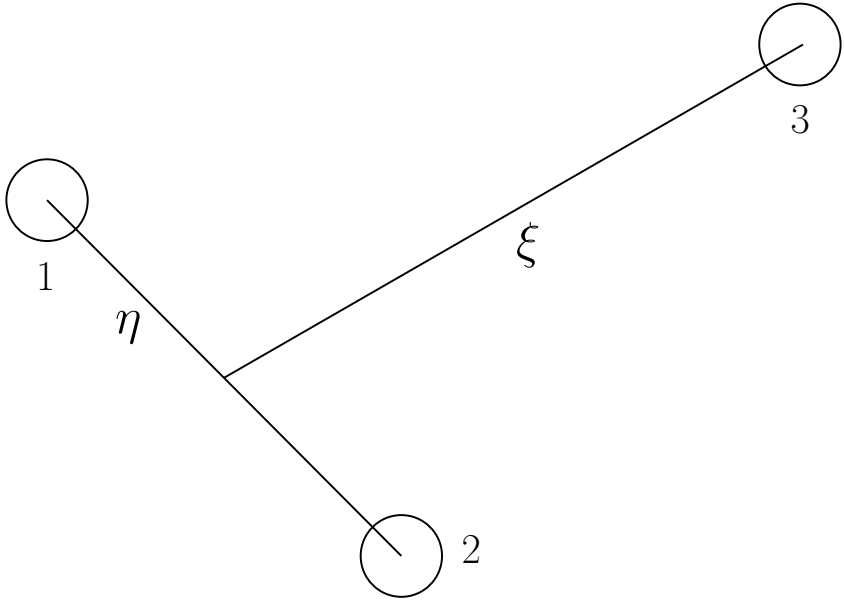,height=8.0cm}}
\caption{Jacobi coordinates }
\end{figure}
\vspace{.20in}
where, of course, the $\vec{r}_{i}$ give us the locations of the 3 particles.
This is a canonical transformation and insures that in the kinetic energy
there are no cross terms.

\esp
The variables $\vec{\xi}$ and $\vec{\eta}$ are
involved separately in the Laplacians and we may consider them as acting in
different spaces. We introduce a higher dimensional vector $\vec{\rho} =
(\ba{c}\vec{\xi} \\ \vec{\eta}\ea)$ and express it in a hyperspherical
coordinate system ($\rho$ and the set of angles $\Omega$).
If we  factor a term of $\rho^{5/2}$ from the solution of the relative
Schr\"{o}dinger equation, i.e. we let $\psi =
\phi/\rho^{5/2}$, we are led to:
\beq
\left[ - \frac{\partial^2}{\partial \rho^2} + H_{\rho} - \frac{2 m E}{\hbar^2}
\right] \phi(k,\rho,\Omega) = 0 ,
\eeq
where
\beq
H_{\rho} = - \frac{1}{\rho^2} \left[ \nabla^2_\Omega - \frac{15}{4} \right] +
\frac{2 m}{\hbar^2} V(\rho, \Omega),
\eeq
$m$ is the mass of each particle, $E$ is the relative energy in the
center of mass and $k^{2}$ is the relative energy multiplied by 
$2m/\hbar^{2}$. $\nabla^2_{\Omega}$ is the purely angular part of the
Laplacian.

\esp
We now introduce the adiabatic basis, which consists
of the eigenfunctions of part of the Hamiltonian: the angular part of the
kinetic energy and the potential :
\beq
H_{\rho} B_{\ell}(\rho,\Omega) = \Lambda_{\ell}(\rho) B_{\ell}(\rho,\Omega),
\label{eigen1}
\eeq
where $\ell$ enumerates the solutions.

\esp
   Using this adiabatic basis, we can now rewrite the
Schr\"{o}dinger equation as a system of coupled
ordinary differential equations.

\esp
\noindent
We write
\beq
\phi(k,\rho,\Omega) = \sum_{\ell^{\prime}} B_{\ell^{\prime}}(\rho,\Omega)
 \, \tilde{\phi}_{\ell^{\prime}} (k,\rho)
\eeq
and obtain the set of coupled equations
\beq
 \left( \frac{d^{2}}{d\rho^{2}} - \Lambda_{\ell}(\rho)  +  k^{2} \right) \,
\tilde{\phi}_{\ell}(k,\rho) + 2\sum_{\ell^{\prime}} C_{\ell,\ell^{\prime}}\,
\frac{d}
{d\rho} \, \tilde{\phi}_{\ell^{\prime}}(k,\rho) 
 +  \sum_{\ell^{\prime}}D_{\ell,\ell^{\prime}}\,
\tilde{\phi}_{\ell^{\prime}}(k,\rho) = 0 ,
\label{eigen2}
\eeq
where we defined:
\beq
C_{\ell,\ell^{\prime}}(\rho)  =  \int d\Omega
\,B^{\ast}_{\ell}
(\Omega,\rho)\frac{\partial}{\partial\rho}B_{\ell^{\prime}}(\Omega,\rho) 
\hlf 
D_{\ell,\ell^{\prime}}(\rho)  =  \int d\Omega \,
B^{\ast}_{\ell}
(\Omega,\rho)\frac{\partial^{2}}{\partial\rho^{2}}B_{\ell^{\prime}}
(\Omega,\rho) .
\ \label{Dmatrix}
\eeq
We note that
\beq 
 D_{\ell,\ell^{\prime}} = \frac{d}{d\rho}\left(C_{\ell,\ell^{\prime}}\right)
 + \left(C^{2}\right)_{\ell,\ell^{\prime}} ,
\label{anti}
\eeq
and that $C$ is antisymmetric and $C^2$ is symmetric.
\newpage
\begin{center}
\section*{Without Bound States}
\end{center}
\vspace{.3cm}

Let us first keep things as simple as possible.
When there are no bound states, we may
write for the relative part of the 3-body trace:
\beq
Tr(e^{-\beta H_3}) = \int \!\! d\vec{\rho} \int \!\! dk \sum_{i}
\psi^{i}(k,\vec{\rho})  (\psi^{i}(k,\vec{\rho}))^{\ast} \;
e^{-\beta (\frac{\hbar^2}{2 m}k^2)} \, ,
\eeq
where we have introduced a complete set of continuum eigenfunctions.
Expanding in the adiabatic basis, we obtain
\beq
Tr(e^{-\beta H_3}) = \int \!\! d{\rho} \int
\!\! dk \sum_{i,\ell} |\tilde{\phi}^{i}_{\ell}(k,\rho)|^2 \;
e^{-\beta (\frac{\hbar^2} {2 m}k^2)} ,
\eeq
where we note that we have integrated over the angles and taken advantage
of the orthogonality of our $B_{l}$'s. We integrate from $0$ to $\infty$.

We now return to our expression for $b_3 - b^0_3$ and proceed as above, 
drop the tildes (also in the subsequent equations), to obtain:
\beq
 \frac{3^{3/2}}{ \lambda_{T}^3} \int \!\! dk\, e^{-\beta E_{k}}
\int \!\! d{\rho}
\sum_{i,\ell}[ (|{\phi}^{i}_{\ell}|^2 - |{\phi}^{i}_{\ell,0}|^2 )
- ( |\bar{\phi}^{i}_{\ell}|^2  -  |\bar{\phi}^{i}_{\ell,0}|^2 )] ,
\eeq
where we have evaluated the trace corresponding to the center of mass.
The amplitudes ${\phi}^{i}_{\ell}$ correspond to
$H_3$, $\bar{\phi}^{i}_{\ell}$ to $H_2 + T_1$ and
amplitudes with a zero belong to the free particles.
%
We now make use of a Wronskian type procedure to evaluate the $\rho$ integrals.
We first write
\beq
 \int_{0}^{\rho_{max} }\!   \sum_{\ell}  |\phi^{i}_{\ell}(k,\rho)|^2 \;
d\rho =
\lim_{k^{\prime} \rightarrow k}  \int_{0}^{\rho_{max}} \!
 \sum_{\ell} \phi^{i}_{\ell}(k,\rho) \phi^{i}_{\ell}(k^{\prime},\rho)
\, d\rho
\eeq
and then, there is the procedure:
\beqa
 \int_{0}^{\rho_{max} }\!   \sum_{\ell} \; (\!\! & \!\phi^{i}_{\ell}(k,\rho)&
\!\! \phi^{i}_{\ell}(k^{\prime},\rho))  d\rho =  \nonumber\\
\frac{1}{k^2 - (k^{\prime})^2} \sum_{\ell}\;[\! & \!\!\phi^{i}_{\ell}(k,\rho)&
\!\! \frac{d}
{d\rho}\phi^{i}_{\ell}(k^{\prime},\rho) - \phi^{i}_{\ell}(k^{\prime},\rho)
\frac{d}{d\rho}\phi^{i}_{\ell}(k,\rho)] ,
\eeqa
evaluated at $\rho = \rho_{max}$. \\
\noindent
---------------------------------------------------------------- \\
I.e. our identity is (see Appendix B):
\beqa
&& \ \sum_{\ell}     \frac{d}{d\rho} \left[\phi_{\ell}(k^{\prime},\rho)
\frac{d}{d\rho}
\phi_{\ell}(k,\rho) - \phi_{\ell}(k,\rho)\frac{d}{d\rho}\phi_{\ell}(k^{\prime},\rho)\right] 
\nonumber \\
&& + \left(k^2 - (k^{\prime})^2\right)\sum_{\ell} \phi_{\ell}(k,\rho) \
\phi_{\ell}(k^{\prime},\rho)  \nonumber \\
&& + 2 \ \sum_{\ell,{\ell}^{\prime}} \frac{d}{d\rho}
\left[ \phi_{\ell}(k^{\prime},\rho)
\ C_{\ell,{\ell}^{\prime}} \ \phi_{{\ell}^{\prime}}(k,\rho) \right] = 0 \label{iden}
\eeqa
and we integrate with respect to $\rho$. Using then the fact that $\phi$
goes to zero, as $\rho$ itself goes to zero, and that C decreases fast
enough for $\rho$ large, we are left with the expression
displayed earlier. \\
---------------------------------------------------------------- \\
\noindent
We now put in the asymptotic form of our oscillatory solutions,
valid for $\rho_{max}$ sufficiently large, given the set of
$\ell$'s, and use l'Hospital's rule to take the
limit as $k^{\prime}  \rightarrow k$. \\
\noindent
The solutions are:
\beq
\phi^{i}_{\ell}(k,\rho) \rightarrow (k\rho)^{1/2}\, {\cal C}_{\ell}^{i}(k) \,
[\cos \delta^{i}(k)\: J_{K+2}(k\rho) - \sin \delta^{i}(k)\: N_{K+2}(k\rho)]
\label{asymp}
\eeq
where the order $K$ is one of the quantities specified by $\ell$.

\esp
Obviously this requires comment. We show, in Appendix A, how given 
N coupled equations, we can choose the solutions so as to diagonalize a 
matrix W, analogous to an R-matrix, but characterising 
adiabatic solutions.  These might be called 
`eigenstates of the standing waves'.  

These will have the property that, for each solution, a unique
eigenphase shift will appear in all of the associated amplitudes.
Further, in the asymptotic regime, the adiabatic functions $B_{\ell}$, 
will - in the absence of interaction - reduce to hyperspherical harmonics,
characterized, in part, by the order $K$. This, in turn implies that in the 
asymptotic form for the amplitude, there appears a term  ${K+2}$. 

The $\cal{C}$ is the coefficient 
of mixture, which tells us how much of each amplitude appears in each 
solution. Of course, we choose orthonormal eigenfunctions.   

\esp
Inserting  (\ref{asymp}) into our integrals we find that
\beq
\sum_{\ell} \int_{0}^{\rho_{max}}\! |\phi^{i}_{\ell}(k,\rho)|^2 \:  d\rho
\rightarrow
\frac{1}{\pi} \frac{d}{dk} \delta^{i}(k)
+ \frac{1}{\pi} \rho_{max} \: + \: osc. \: terms
\eeq
and, thus, that
\beq
\int_{0}^{\rho_{max}}\! (| \phi^{i}_{\ell}(k,\rho)|^{2}
 -  |\phi^{i}_{\ell,0}(k,\rho)|^2) \; d\rho  \rightarrow
\frac{1}{\pi} \frac{d}{dk} \delta^{i}(k) \: + \: osc. \: terms
\eeq
We let $\rho_{max}$ go to infinity, and the oscillatory terms
- of the form $\sin(2 k \rho_{max} + \cdots)$ - will
not contribute to the subsequent integration over $k$.
[This is in 3 dimensions. In one dimension, we have shown that even for the 
second virial coefficient\cite{all}, the oscillatory terms can give a 
contribution.] 
Our basic formula now reads:
\beq
b_3 - b^0_3 \, = \, \frac{3^{3/2}}{\pi\lambda^3_{T}}\int_{0}^{\infty}
\! dk \;\frac{d}{dk} \, G(k) \;
e^{-\beta \frac{\hbar^2}{2 m} k^{2}}
\eeq
where
\beq
G(k) = \sum_{i}\:  \delta^{i}(k) -  \sum_{j}\: \bar{\delta}^{j}(k)
\eeq
The first $\delta's$ arise from comparing three interacting particles
with three free particles. The second $\bar{\delta's}$  arise when a 3-body
system, where only two particles are interacting (one particle being a
spectator), is compared to three free particles.

\esp
The evaluation of these sums (and their difference) is a delicate matter.
The individual sums are separately infinite, as they correspond, classically
to parts of the three-body cluster which depend on the volume.
It is crucial to subtract 
individual, respective terms - before summing - to obtain a finite result.

We also observe that, if we limit ourselves to a finite number of the coupled 
equations, there will be fewer of the $\delta's$ than of the  $\bar{\delta's}$,
due to the requirements of symmetry,  but that the $\delta's$ will, in their
totality, be as important as the $\bar{\delta's}$ because they are 
associated with 3 binary potentials, while the
$\bar{\delta's}$ will only reflect the potential of a pair.

\newpage
\begin{center}
\section*{Bound States}
\end{center}
\vspace{.3cm}

Of course, we shall need again to evaluate the 
relative part of the Bose trace: 
\begin{equation}
Tr[(e^{-\beta H_3} - e^{-\beta T_3})
-  \, (e^{-\beta(H_2 + T_1)} - e^{-\beta (T_2 + T_1)})] 
\end{equation}
We shall have to discuss the contribution of each of the terms above but 
(as before) only integrating
over a radius $\rho$ extending to an appropriately large value of $\rho_{max}$. The terms in
$\rho_{max}$ must cancel.

\esp
If there are bound states, the major change in the eigenpotentials,
see Eqs. (\ref{eigen1}, \ref{eigen2}), 
is that for some of these potentials, the $\Lambda's$, instead of going to 
zero at large distances (large $\rho$), tend to a negative `plateau'. 
I.e., the eigenpotentials become flat and negative. 
We illustrate this with an example from our work in one dimension,
with attractive delta-function potentials\cite{alp}. 

  Here we present, illustrate, the dominant diagonal part of the adiabatic
coupling matrix for the first few equations. In Fig. 2, we plot the
first few even eigenvalues $\Lambda_{K}$ (with $K = 0,6,12, ...$)
as a function of $\rho$.

\begin{figure}[h]
\input{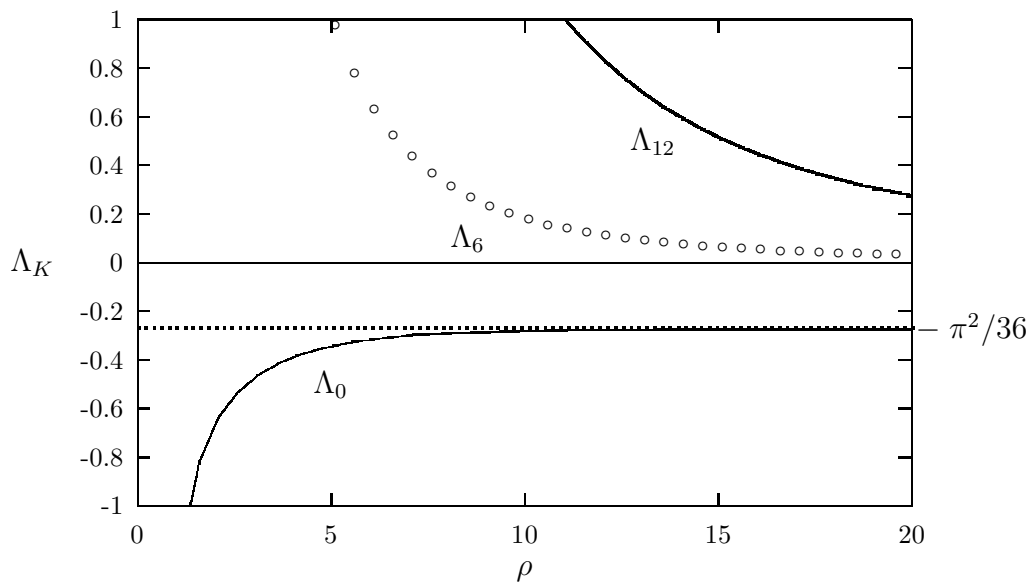}
\caption{Adiabatic eigenvalues $\Lambda_K$ for $c = -1$ (see Eq. \ref{para}).
 The points
 represent the two-body bound state energy level at $-\pi^{2}/36$}.
\end{figure}

The plateau is the indication that, for the amplitude associated with
that particular $\Lambda$, asymptotically the physical
system consists of a 2-body bound state and a free particle.
In the case above, the eigenpotential also `supports' one 3-body bound state.

\newpage

\esp
More generally the eigenfunction expansion of the trace associated 
with $H_3$, will read as follows:
\beqa
\smn &\mbox{}& Tr(e^{-\beta H_3}) =   \nonumber \\
\smn &\mbox{}& \sum_{m} e^{-\beta E_{3,m}} + 
\sum_{i} \int_{0}^{\infty} \!\! dq_0 \!
\int \!\! d\vec{\rho}\,  \psi^{i}(q_0,\vec{\rho})\,
(\psi^{i}(q_0,\vec{\rho}))^{\ast} \;
e^{-\beta (\frac{\hbar^2}{2 m}q_0^2 -\epsilon_{2,0})}    
\label{expan}
\eeqa
where we have used the wavenumber $q_0$ to allow us to include the 
contribution for $E < 0$ from solutions
which have amplitudes which are still oscillatory for negative energies 
(above that of the deepest 2-body bound state). 
The contribution from these solutions when $E > 0$ is, of course, still
included. \\
For present and later purposes, we define 
$q_j$'s by $k^2 = q_j^2 - (2m/\hbar^2)\epsilon_{2,j}$, where
$\epsilon_{2,j}$ is the binding energy of the corresponding 2-body bound state.
The limit $\bar{q}_{j}$ equals  $\sqrt{\frac{2 m}{\hbar^2}\epsilon_{2,j}}$.

\subsection*{Adiabatic Approximation - One 2-body Bound State}
Assume, now, that we have one 3-body bound state and, in addition, one
2-body bound state, associated with both
one eigenpotential for $H_3$ and one for $H_2 + T_1$, 
and introduce amplitudes.
\footnote{As we will see in the next section, we can have more than
one eigenpotential associated with a particular 2-body bound state, 
even for the same Hamiltonian.}
Assume further that we neglect the coupling between the differential 
equations, thus resorting to an adiabatic approximation. Since each 
eigenstate is now associated with one amplitude $\phi^{i}_{i}$,
we will write it as $\phi^{i}$. Further, the ${\cal C}'s$, which arise 
from the diagonalization of the $W$- matrix, and which state how the 
eigenstates of the W-matrix are composed of the different
amplitudes, and which give rise to the `mixture coefficients', can now 
be set to $1$ for each of the `diagonal amplitudes'. It is also useful here
to keep two variables: our old $k$ and a $q$.

\esp
\noindent
For $H_3$, the asymptotic behaviour will be as follows.  

\esp
\esp
\noindent
For $E > 0$. \\
For $i > i_{0}$, where $i_{0}$ stands for the lowest value of the index $i$, 
associated with the adiabatic eigenfunction $B_{0}$, corresponding to 
the lowest eigenpotential, and for the value $i_0$:
\beq
\phi^{i}(k,\rho) \rightarrow (k\rho)^{1/2}\,
[\cos \delta^{i}(k) \, J_{K_{i}+2}(k\rho) - \sin \delta^{i}(k) \,
N_{K_{i}+2}(k\rho)]
\eeq
\beq
\phi^{i_0}(k,\rho) \rightarrow (k\rho)^{1/2}\,
[\cos \delta^{i_0}(k) \, J_{K_{i_{0}}+2}(q\rho) - 
\sin \delta^{i_0}(k) \, N_{K_{i_{0}}+2}(q\rho)]  
\eeq
See Larsen and Poll\cite{lap} and Appendix A. 
We have also written $q_{i_0}$ as $q$. 

\esp
\esp
\noindent
Using our procedure as before, for each $\phi^{i}$ ($i \neq i_0$) 
we obtain for the integral over $\rho$:
\beq
\frac{1}{\pi}\frac{d}{dk} \delta^{i}(k) + \frac{\rho_{max}}{\pi}
\eeq 
and for the case $i_0$, we find 
\beq
\frac{1}{\pi}\frac{d}{dk} \delta^{i_0}(k) + 
\left(\frac{k}{q}\right)\frac{\rho_{max}}{\pi}
\eeq

\esp
\noindent
For $E < 0$.
\beq
\phi^{i_0}(q,\rho) \rightarrow (q\rho)^{1/2}
[\cos \delta^{i_o}(q) \, J_{K_{i_{0}}+2}(q\rho) - \sin \delta^{i_0}(q) \,
N_{K_{i_{0}}+2}(q\rho)]
\eeq
which then yields
\beq
\frac{1}{\pi}\frac{d}{dq} \delta^{i_0}(q) + \frac{\rho_{max}}{\pi}
\eeq
\esp
We note that in the expressions above,  
 $\phi^{i_0}(k,\rho)$ and $\phi^{i_0}(q,\rho)$,
have different functional dependences on their arguments and, of course,
different normalization factors, due
to the different integrations, respectively over $k$ and $q$.

\esk
\noindent
{\bf Let us now consider the term $(H_2 + T_1)$.} \\

\esp
The discussion absolutely parallels that for $H_3$.  \\

\esp
The eigenpotentials are perfectly similar to those of $H_3$. See Figure 3.
\esp
\esp
\esp
\esp
\esp
\esp
\esp
\esp
\begin{figure}[h]
\centerline{\psfig{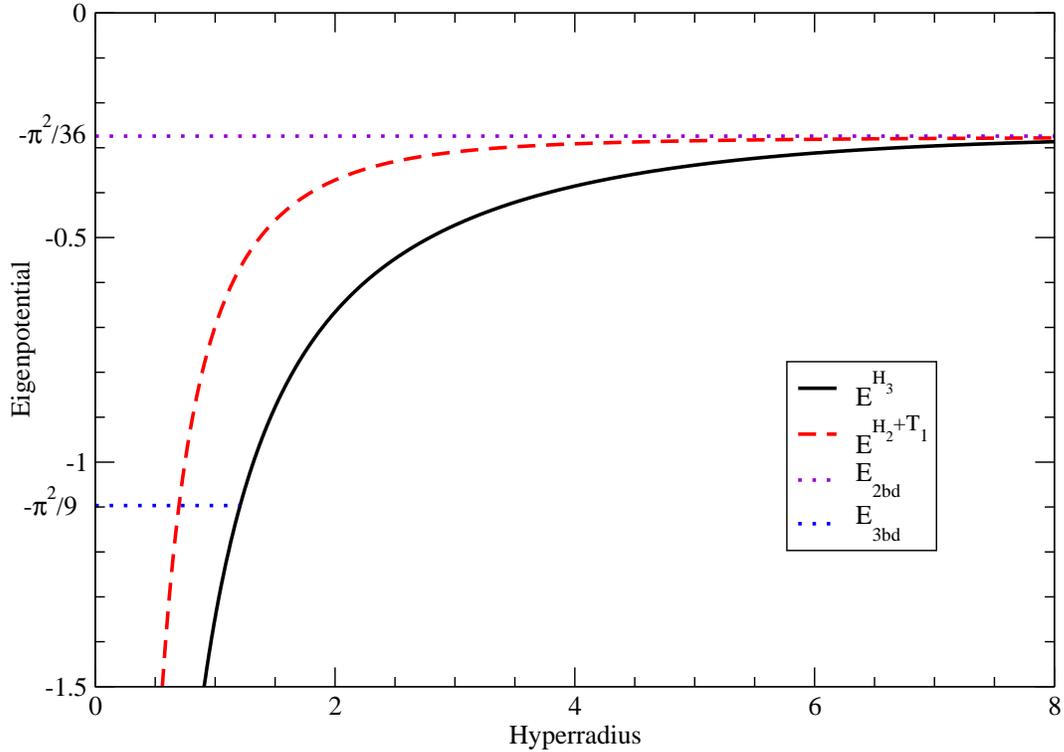}}
\esp
\esp
\esp
\esp
\caption{Eigenpotentials for $H_3$ and $H_2 + T_1$}
\end{figure}
\esp
Using the index $j$ instead of $i$, but in the same manner,
we note that 
the attractive part of the eigenpotential associated with
$\overline{B}_{j_0}$ is simply weaker that 
of $B_{i_0}$, and will not sustain a 3-body bound state.
The eigenpotential potential will however tend, for large values 
of $\rho$, to the same fixed negative energy, which
is that of the 2-body bound state.

\esp
\esp
\noindent

\esp
\esp
\noindent
{\bf Finally for $T_3$ and for $(T_2 + T_1)$} , obviously there are only 
solutions for $E > 0$.

\esp
Since they appear for the same set of equations as considered
with interactions, but now without these interactions, we can 
associate them with the previous indices $i$ and $j$. The respective
$B$'s and $\overline{B}$'s reduce to the corresponding hyperspherical
harmonics.

\subsection*{Putting it all together ...}

For $E < 0$, then, we see that we can subtract the contribution of 
the element stemming from $H_3$, 
from the corresponding element stemming from  $(H_2 + T_1)$, 
and obtain the result 
\beq
\frac{1}{\pi} \frac{d}{dq}[\delta^{i_0}(q) -  \bar{\delta}^{j_0}(q)] 
\eeq
Essential is the fact that the coefficient of $\rho_{max}$ cancels ! \\

\noindent
For $E > 0$, and $i$ and $j$ greater than $i_0$ and $j_0$, respectively,
the terms deriving from $H_3$ and $T_3$ subtract, so as to eliminate 
the $\rho_{max}$ contribution, and so do the terms from $(H_2 + T_1)$ 
and $(T_2 + T_1)$.
\\
We therefore obtain results such as:
\beq
\frac{1}{\pi} \frac{d}{dk}[\delta^{i}(k) -  \bar{\delta}^{j}(k)], 
\eeq
where again again we can associate $i$ and $j$, and the $\rho_{max}$ 
contribution cancels. \\

For $i_0$ and $j_0$, the results from $H_3$ and  $(H_2 + T_1)$ 
subtract, thereby eliminating the $\rho_{max}$ dependence. (See, below.)

\esk
{\bf  
In the adiabatic approximation, we can then write the following formula 
for the relative part of the complete trace, in the case of one 3-body 
bound-state + one 2-body bound state, associated with one eigenpotential:}
\esp
\beqa
\smn &\mbox{}&
Tr[(e^{-\beta H_3} - e^{-\beta T_3}) 
-  \, (e^{-\beta(H_2 + T_1)} - e^{-\beta(T_2 + T_1)} )] =  \nonumber \\
\esp \esp \esp
\smn &\mbox{}&  \,  e^{-\beta E_{3}} + 
\frac{1}{\pi}  \int_{0}^{\infty} \!\! dk \!
\, \frac{d}{dk} [\sum_{i > i_0}\delta^{i}(k) - \sum_{j>j_0} \bar{\delta}^{j}(k)]
\, e^{-\beta (\frac{\hbar^2}{2 m}k^2)} \nonumber \\
\smn &\mbox{}&
\smp  + \,  \frac{1}{\pi} \: e^{\beta \, \epsilon_{i_0}} 
\int_{0}^{\infty}  \!\! dq \!
\, \frac{d}{dq} [\delta^{i_0}(q) -  \bar{\delta}^{j_0}(q)]
\, e^{-\beta (\frac{\hbar^2}{2 m}q^2)} .
\eeqa

Alternatively, the last integral could be split into two, 
one from $0$ to $\overline{q}_{i_0}$, integrating over $q$, together 
with one from $\overline{q}_{i_0}$ to $\infty$, integrating over $k$.
That is:

\esp
\beqa
\smn &\mbox{}&
Tr[(e^{-\beta H_3} - e^{-\beta T_3})
-  \, (e^{-\beta(H_2 + T_1)} - e^{-\beta(T_2 + T_1)} )] =  \nonumber \\
\esp \esp \esp
\smn &\mbox{}&  \,  e^{-\beta E_{3}} +
\frac{1}{\pi}  \int_{0}^{\infty} \!\! dk \!
\, \frac{d}{dk} [\sum_{all \, i}\delta^{i}(k) - 
\sum_{all \, j} \bar{\delta}^{j}(k)]
\, e^{-\beta (\frac{\hbar^2}{2 m}k^2)} \nonumber \\
\smn &\mbox{}&
\smp  + \,  \frac{1}{\pi} \: e^{\beta \, \epsilon_{i_0}}
\int_{0}^{\overline{q}_{i_0}}  \!\! dq \!
\, \frac{d}{dq} [\delta^{i_0}(q) -  \bar{\delta}^{j_0}(q)]
\, e^{-\beta (\frac{\hbar^2}{2 m}q^2)} .
\label{two}
\eeqa

\esp
For {\bf simplicity}, at this stage, we do not present formulas for 
the cases involving more bound states and eigenpotentials. 
They would be similar to the ones that we have just presented but involve 
more indices.
[In the next section, we will consider a case with 4 eigenpotentials,
with the same asymptotic behaviour, the same eigen-energy, 
and therefore the same $\epsilon_{i_{0}}$.  We will sum the contributions 
of the 4 states.] 

\esp
In another remark we note that in order for the $\rho_{max}$ 
contributions for
$E < 0$ to cancel, we need a one-to-one correspondence
between the `bound-state' eigenpotentials for $H_3$ and those 
for $H_2 + T_1$. This will 
also be important in the results for $E > 0$, 
when these eigenpotentials are involved.
\newpage
\begin{center}
\section*{McGuire's Model With Attractive Forces}
\end{center}
\vspace{.3cm}
To show concrete details, and since to our knowledge there are 
no published results for this important case, we revisit our
previous work with McGuire's
model of three particles in one dimension\cite{asm}, 
but this time letting the delta function interactions be attractive.
This would also be a prime example of the usefulness of adiabatic 
approximations.

A salient aspect of this model, with identical masses and interactions,
is that there is no "breakup", i.e. there is no possibility of a system,
characterized asymptotically by two bound particles and a free one,
to evolve to a system characterized asymptotically 
by three free particles. To quote McGuire\cite{mcg}: "there are no new 
velocities generated even though there are three particles present".

Similarly, when one considers the case of one particle that does not
interact with the other two, which do interact but are of equal masses,
there is no possibility of generating new velocities. 

\esp
Technically, this manifests itself in that the R-matrix, associated with the
S-matrix that characterizes the 
scattering of the 3 bodies, has off diagonal elements which are zero, when 
linking solutions of the two types of physical outcome mentioned 
above\cite{chuka}. In our adiabatic-basis formalism our matrix $W$, related
to the R-matrix by $R = - W^{-1}$, behaves similarly in the relevant 
off-diagonal elements. In our present work, we force the issue by using
adiabatic approximations, thus insuring that 
solutions which have asymptotic forms 
associated with eigenpotentials that tend to zero for large $\rho$, 
and have non-zero amplitudes only for $E > 0$,  do NOT couple with the
solutions associated with eigenpotentials
that tend to a negative fixed-energy when $\rho$ becomes large.

\esp
We find then, thanks to this dichotomy, that the better part
of our calculation is already done!
The part, associated with the $\Lambda$'s that go to zero for large
$\rho$, has already been calculated to first order 
in our previous article where no bound states appear;
we need only an overall change in sign.

The expansions for the $q_K$'s that started with $(K+3)$ or $(K+1)$
now start with $(K-3)$ or $(K-1)$. The expressions for the $C$'s 
or the $D$'s now involve $(K-3)$ or $(K-1)$. 
The sums that used to start with $K=0$, $K=3$ for $H_3$ and $K=0$, $K=1$
for $H_2 + T_1$ now, in the attractive case, start with  $K=6$, 
$K=9$ for $H_3$ and $K=2$, $K=3$ for $H_2 + T_1$. 

Thus, to first order in $k$, the wave number, the sums and differences
of the phase shifts (and their derivatives) remain identical to the 
results that we have previously achieved, apart from an overall 
change in sign. The contribution to the fugacity coefficient - which 
is additive to the other contributions that we must still 
calculate - also just changes in its sign.

Our result SO FAR is then:
\begin{equation}
b_{3}-b_{3}^{0}=+\frac{3\sqrt{2}}{4\pi g}\,\frac{1}{\beta} \label{fresult}
\end{equation}
where $\beta$ is the same variable that appears in Eq.(1), associated with 
the inverse of the temperature, and $g$ is the `strength' of the delta
functions in our model.
\footnote{Please note that a factor of 2 is missing from the equation 
following Eq.(78) in our previous paper. This is clear from the previous 
two lines.}

[We note that a factor of $(-1)^m$ is missing from Eqs.(13) and (15) of
our previous article.
The sine versions of our adiabatic bases should read: \\
For $H_3$ and 
$K=\textrm{odd}, \, \, \, A_{K}(\rho)\, \,(-1)^m \, \sin(q_{K}(\vartheta-\frac{m\pi}{3})) $ \\
For $H_2+T_1$ and 
$K=\textrm{odd}, \, \, \, A^{12,3}_{K}(\rho)\, \,(-1)^m \, \sin(q_{K}(\vartheta-m\pi)) $ ]

\esp
\esp
\subsection*{The new derivative terms}
We mentioned that in the repulsive case the sums used to start with $K=0$, 
$K=3$ for $H_3$ and $K=0$, $K=1$ for $H_2 + T_1$.  
In the attractive case, each of these cases is associated 
with a different eigenpotential,
which however tends to a common plateau of energy $- \pi^2/36$, reflecting 
the asymptotic behaviour of the products:
two bound particles (each time with the same binding energy) and a free one.

Below, we illustrate the behaviour of the eigenpotentials, adding the 
diagonal part of the adiabatic coupling elements, see Figure 4.
We also show the linear behaviour 
of the phase shifts (calculated numerically) for small values of the $q_K$'s.
See Figure 5.

\esp
\vspace{.9cm}
\begin{figure}[h]
\centerline{\psfig{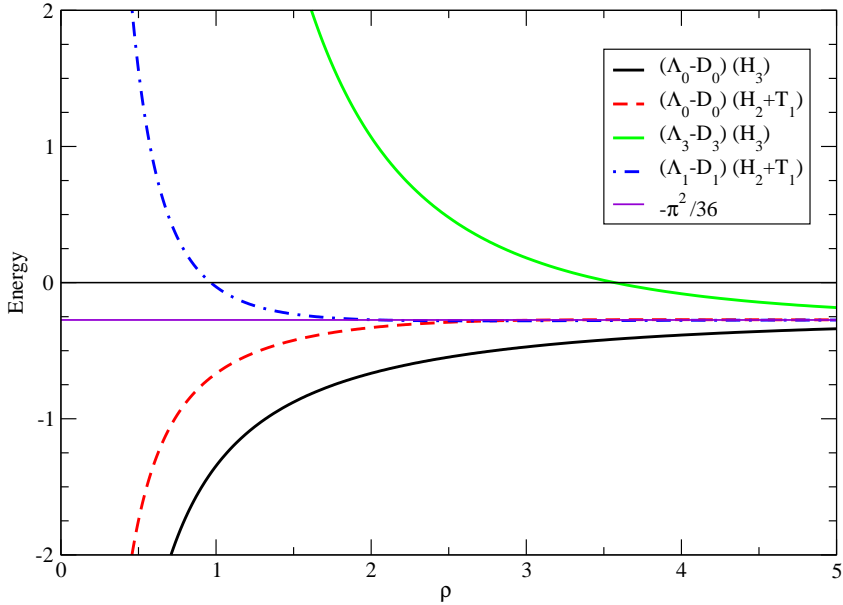}}
\esp
\esp
\esp
\caption{The diagonal part of the interactions for the four cases.}
\end{figure}

We note that to be consistent in our work,
we only need the slope of the phases near the origin.
For the case of $H_3$ and $K = 0$, we will use the slope of the exact
phase shift, as it is known and is associated with a zero energy resonance.
See Appendix D.

\esp
\vspace{.9cm}
\begin{figure}[h]
\centerline{\psfig{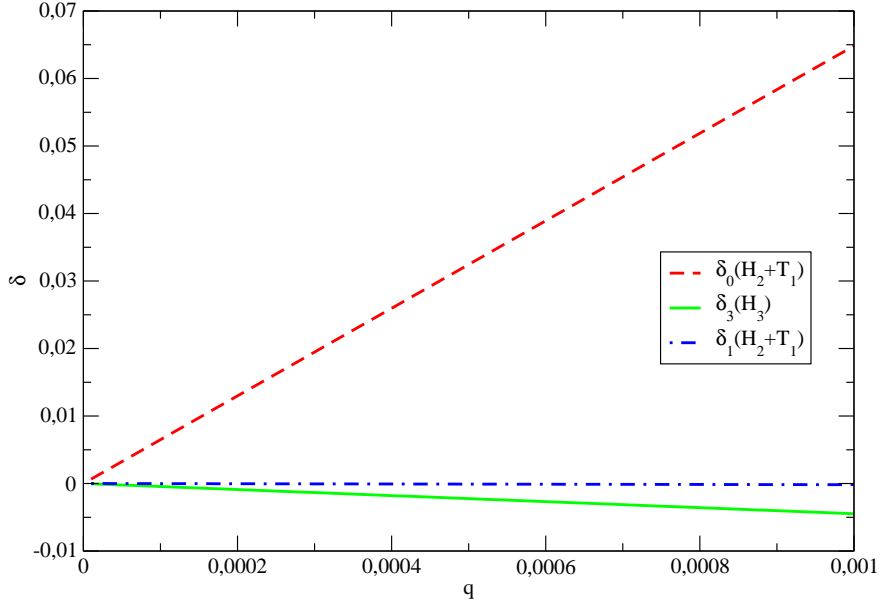}}
\esp
\esp
\esp
\caption{The low q behaviour of the phase shifts.}
\end{figure}

Taking the slopes of the phase shifts - and trying to round and keep 
significant digits -  we find:  \\
for $H_3$: $-8 \sqrt{3} /\pi$, $-4.47$ for $K= 0$ and $K = 3$ respectively, 
and \\
for $H_2 + T_1$: $64.91$ and $-.176$ for $K = 0$ and $K = 1$,  \\
we obtain for the derivative of the sum of the $\delta$'s minus the sum
of the $\overline{\delta}$'s: $-73.62$, of course rounding and trying to
keep meaningful numerical results. 

The bound state contribution in Eq.(35) now reads: 
\beq
 e^{\beta \frac{\hbar^2}{2 m}\frac{\pi^2}{9}} 
 +  \frac{1}{\pi} \! e^{\beta \, \frac{\hbar^2}{2 m} \, \frac{\pi^2}{36}} \, \, 
\frac{\sqrt{\pi}}{2} (-73.62) \sqrt{\frac{2 m}{\beta \hbar^2}}
\eeq

[To obtain the extra contribution to the fugacity coefficient, we need to 
multiply our result (above) by $\frac{\sqrt{3}}{ \lambda_T}$.
We also need to look at the  
contributions of `new' oscillatory terms.]

\subsection*{The new oscillatory terms} 
Our work, here, parallels that which led to expression (34) in our previous
paper. \\
\esp
For each of our four cases, $K = 0, 0, 1, 3 \:$ we have to consider 
\beq
\frac{1}{q} \, \sin(2 q \rho_{max} + 2 \delta^K) \; \;
{\rm and} \; \;  \frac{1}{k}\,\sin(2 k \rho_{max} - K \pi
+ \pi/2) 
\label{osc}
\eeq
where the first term is associated with an interaction which, in each case,
asymptotically does not have any $1/\rho^2$ dependence, and must be integrated
over $q$.
The second term  is a `no-interaction'  term, and must be integrated
over $k$..

Let us consider  the  free or 'no-interaction' terms in 
(\ref{osc}) namely  
\begin{equation}
\frac{1}{k}  \sin\left(2 k \rho_{max} - K \pi + \frac{\pi}{2} \right)  \ ,
\end{equation}
 $K=0,0,1,3$.
In the past (no bound states) we have subtracted them directly  
from the interaction terms.. Here we cannot do so.
However, since the number of terms considered in  (\ref{osc}) is  finite, 
the terms involved in the sum can be reordered and 
yield:
\beqa
 -\frac{1}{k}  \sin\left(2 k \rho_{max} + \frac{\pi}{2} \right)  + 
\frac{1}{k}  \sin\left(2 k \rho_{max} + \frac{\pi}{2} \right)  \nonumber \\
 -\frac{1}{k}  \sin\left(2 k \rho_{max} - 3 \pi + \frac{\pi}{2} \right)  
+ \frac{1}{k}  \sin\left(2 k \rho_{max} -  
\pi + \frac{\pi}{2} \right) =0 \nonumber
\eeqa
We now consider the interaction terms 
(\ref{osc}), namely
\begin{equation} 
\frac{1}{q}  \sin\left(2 q \rho_{max} + 2 \delta^K\right)  \;
{\rm and} \;  \, \frac{1}{q}  \sin\left(2 q \rho_{max} + 
2 {\bar \delta}^K\right) 
\end{equation}
for K=0,3  and K=0,1 respectively. \\
We recall that we call $\delta$'s the phases for $H_3$, and 
${\bar \delta}$'s for $H_2 + T_1$.
\esp
\esp

These phases go to zero as $q \rightarrow 0$, except for 
the symmetric case $K=0$, where $\delta  =3 \pi/2$.
This is important because in the resulting integrals, upon the change of
variables to $y$, {\it see below},  each  $\delta$ will be 
evaluated as $\delta( y/{2 q \rho_{max}})$, in the limit of $\rho_{max} 
\rightarrow \infty$. 
This implies that the expression
\begin{eqnarray}
&& \frac{1}{q}  \sin (2 q \rho_{max} + 2 \delta^0)-  
\frac{1}{q}  \sin (2 q \rho_{max} + 2 {\bar \delta}^0) \nonumber\\
&+&\frac{1}{q}  \sin (2 q \rho_{max} + 2 \delta^3)-  
\frac{1}{q}  \sin (2 q \rho_{max} + 2 {\bar \delta}^1) \nonumber
\end{eqnarray}
 is equal to  
\begin{equation}
-2 \frac{1}{q}  \sin\left(2 q \rho_{max} \right) \ .
\end{equation}
The contribution of the oscillatory term to the final integral 
then reads
\begin{equation}
\frac{\sqrt{3}}{\lambda_T} \int_0^{\infty} dq \frac{1}{2 \pi} 
\left[ -2 \frac{1}{q}  \sin (2 q \rho_{max} ) \right]  
\exp\left(-\beta \frac{\hbar^2}{2 m} \left[q^2 - \frac{\pi^2}{36} \right]\right) 
\end{equation}
Setting $y=2 q \rho_{max}$ we have
\begin{equation}
I=-\frac{\sqrt{3}}{\lambda_T} (\frac{1}{\pi})\, \int_0^{\infty} dy  
\frac{\sin y}{y}  \exp\left(-\beta  \frac{\hbar^2}{2 m}\left[\left(\frac{y}{2 \rho_{max}}\right)^2 -  \frac{\pi^2}{36} \right]\right)
\end{equation}
which, at the limit $ \rho_{max}$ infinite, yields
\begin{equation}
I=-\frac{\sqrt{3}}{2 \lambda_T} \, e^{\beta (\hbar^2/2 m)(\pi^2/36)}
\end{equation}

\subsection*{Our result} 
Recalling that: 
\beq
c =  (2m/\hbar^2)(3g/\pi\sqrt{2}),\ \ g = (\hbar^2/2 m)(\pi\sqrt{2} c/3),  
\ \ \lambda_{T} = (2 \pi \hbar^2 \beta/m)^{1/2} \label{para}, 
\eeq
and that the following result is for the particular value of $c=-1$, we obtain:
\beq
b_3 - b_3^0 = \frac{\sqrt{3}}{\lambda_T} \left(e^{\lambda_T^2 (\pi/36)} 
- \frac{1}{2}\, e^{\lambda_T^2 (\pi/144)} \right) + 
\frac{1}{\lambda_T^2} \left(-\frac{9}{\pi} + 
\sqrt{3} \, (- 73.62) \,
e^{\lambda_T^2 (\pi/144)}\right) \, .
\eeq
We note that the energies in the exponentials correspond to
the energy of the three-body bound state: $-(\hbar^2/2 m) (\pi^2/9) \,c^2$, 
and that of
the two-body bound state: \\$-(\hbar^2/2 m) (\pi^2/36) \,c^2$. 
The numerical value 
$(- 73.62)$ is, of course, a result that has been rounded.  

\newpage
\begin{center}
\section*{The Full Generalization}
\end{center}
{\bf Assume now that for $H_3$} we have one 3-body bound state and 
(asymptotically) two possibilities of a 2-body bound state, 
each corresponding to one eigenpotential, and introduce amplitudes. 
This will be the simplest example that reveals the 
details that we must deal with in general. 
The asymptotic behaviour will be as follows. \\
The upper index $i$ will identify the overall solution, corresponding to
$\psi^{i}$ in Eq.(\ref{expan}). The $\ell$'s will denote the amplitudes.
For us, in our example, $\ell_0$  is associated with the adiabatic 
eigenfunction 
$B_{\ell_0}$, the lowest eigenpotential, 
and $B_{\ell_1}$ with the second lowest. 
Letting also, to simplify the notation, $q= q_0= q_{\ell_0}$ !\\ 
\\
For $E > 0$ and $\ell > \ell_0, \ell_1$, and for the values 
$\ell_0, \ell_1$ we have then: 
\beq
\! \! \! \! \!  \! \! \! \!
\phi^{i}_{\ell}(q,\rho) \rightarrow (q\rho)^{1/2}{\cal C}_{\ell}^{i}(q)\,
[\cos \delta^{i}(q) \, J_{K_{\ell}+2}(k\rho) - \sin \delta^{i}(q) \,
N_{K_{\ell}+2}(k\rho)]
\eeq
\beq
\;
\phi^{i}_{\ell_{1}}(q,\rho) \rightarrow (q\rho)^{1/2}{\cal C}_{\ell_{1}}^{i}(q) 
\, [\cos \delta^{i}(q) \, J_{K_{\ell_{1}}+2}(q_{\ell_1}\rho) - 
\sin \delta^{i}(q) \, N_{K_{\ell_{1}}+2}(q_{\ell_1}\rho)]
\eeq
\beq
\;
\phi^{i}_{\ell_{0}}(q,\rho) \rightarrow (q\rho)^{1/2}{\cal C}_{\ell_{0}}^{i}(q) 
\, [\cos \delta^{i}(q) \, J_{K_{\ell_{0}}+2}(q\rho) - 
\sin \delta^{i}(q) \, N_{K_{\ell_{0}}+2}(q\rho)]
\eeq 
\esp
\esp
\esp
If we then use our procedure as before, summing over the
contributions of all of the amplitudes 
corresponding to the solution $i$, we obtain for the integral over $\rho$:
\beq
\frac{1}{\pi}\frac{d}{dq} \delta^{i}(q) + \frac{\rho_{max}}{\pi}\left(
\sum_{\ell \neq \ell_{0}, \ell_{1}} |{\cal C}_{\ell}^{i}(q)|^2 \frac{q}{k} 
+ |{\cal C}_{\ell_{1}}^{i}(q)|^2 \frac{q}{q_{\ell_1}}
+ |{\cal C}_{\ell_{0}}^{i}(q)|^2 \right)
\label{result1}
\eeq
\esp
To obtain the first term, we have use the fact that  $\cal C$
is an orthogonal matrix, and that therefore 
$\sum_{\ell} |{\cal C}^i_{\ell}|^2 = 1$ .
\\
\\
For $E < 0$, we have to consider two possibilities: 
$0 < E < -\epsilon_{\ell_1}$ and 
$ -\epsilon_{\ell_1} < E  < -\epsilon_{\ell_0}$, where $-\epsilon_{\ell_1}$
and $-\epsilon_{\ell_0}$ are the respective energies of the two 2-body
bound states, the ${\ell_0}$ bound state being the deepest.
We will then have: \\
\\
For $0 < E < -\epsilon_{\ell_1}$, 
\beq
\phi^{i}_{\ell_{1}}(q,\rho) \rightarrow (q\rho)^{1/2} 
{\cal C}^{i}_{\ell_{1}}(q)
[\cos \delta^{i}(q) \, J_{K_{\ell_{1}}+2}(q_{\ell_1}\rho) - 
\sin \delta^{i}(q) \, N_{K_{\ell_{1}}+2}(q_{\ell_1}\rho)]
\eeq
\beq
\phi^{i}_{\ell_{0}}(q,\rho) \rightarrow (q\rho)^{1/2}
{\cal C}^{i}_{\ell_{0}}(q)
[\cos \delta^{i}(q) \, J_{K_{\ell_{0}}+2}(q\rho) - 
\sin \delta^{i}(q) \, N_{K_{\ell_{0}}+2}(q\rho)]
\eeq
which then yields
\beq
\frac{1}{\pi}\frac{d}{dq} \delta^{i}(q) + 
\frac{\rho_{max}}{\pi} \left(
 |{\cal C}_{\ell_{1}}^{i}(q)|^2 \frac{q}{q_{\ell_1}}
+ |{\cal C}_{\ell_{0}}^{i}(q)|^2 \right)
\eeq

Finally, for $ -\epsilon_{\ell_1} < E  < -\epsilon_{\ell_0}$,
we will have 
\beq
\phi^{i}_{\ell_{0}}(q,\rho) \rightarrow (q\rho)^{1/2}
[\cos \delta^{i}(q) \, J_{K_{\ell_{0}}+2}(q\rho) -
\sin \delta^{i}(q) \, N_{K_{\ell_{0}}+2}(q\rho)]
\eeq
which gives
\beq
\frac{1}{\pi}\frac{d}{dq} \delta^{i}(q) +
\frac{\rho_{max}}{\pi} 
\eeq
\newpage
\noindent
{\bf For $(H_2 + T_1)$, the discussions} and the expressions 
follow in a similar fashion. 
\esp
\esp
\\ 
We will change $i$ to $j$, $\ell$ to $r$. 
We will let $q = q_0 = q_{{\ell}_0} = q_{r_0}$. \\
It is clear, since 
the reference is to the energy associated with the 2-body bound state,
that the deepest (lowest) energy of the $H_2 + T_1$ system will 
be same as for the $H_3$ system. The range of  the wave 
number integration in the two cases, will be the same.
For the equivalent of Eq.(\ref{result1}), we will have: 
\beq
\frac{1}{\pi}\frac{d}{dq} \bar{\delta}^{j}(q) + 
\frac{\rho_{max}}{\pi}\left(
\sum_{r\neq {r_0}, {r}_{1}} |{\cal (C')}_{r}^{j}(q)|^2 
\frac{q}{k}
+ |{\cal (C')}_{r_1}^{j}(q)|^2 \frac{q}{q_{r_1}}
+ |{\cal (C')}_{r_0}^{j}(q)|^2 \right)
\label{result2}
\eeq
where $q_{r_1}$ will be equal $q_{\ell_1}$, the energy of the second 
2-body bound state being equal in the two cases! \\

{\bf If we look to subtract} the 
$\rho_{max}$ terms for $H_3$ from those
of $H_2 + T_1$, we note the following.
First, while in our McGuire example there were as many eigenpotentials 
associated with $H_3$ as with $H_2 + T_1$, we don't see that this 
would necessarily apply in the general case. 
Second, {\bf even} if this were to be the case, if we compare the 
$\cal{C}$'s and their factors in the two situations, 
we see no reason why the $\rho_{max}$ terms should cancel. 
Of course, they would do so in the adiabatic case. \\

\noindent
{\bf The situation is now clear.} \\

We have a simple formula that we can use in general, which involves
the energies of the appropriate 3-body bound states and integrals over 
sums of the derivative of respective phase shifts, 
in addition to $\rho_{max}$ terms. 
We shall need to discuss the latter in the next section.

The volume independent result for the 3-body fugacity cluster,
is then given by:
\esp
\beq
b_3 - b_0 =  \frac{3^{3/2}}{\lambda^3_T} \left(
\sum_{m} e^{-\beta E_{3,m}} +
\frac{1}{\pi}  \int_{0}^{\infty} \!\! dq \!
\, \left[\sum_{all \, i} \frac{d}{dq} \delta^{i}(q) - 
\sum_{all \, j} \frac{d}{dq} \bar{\delta}^{j}(q)\right]
\, e^{-\beta (\frac{\hbar^2}{2 m}q^2-\epsilon_{2,0})} \right)
\eeq
\esp

A comment about oscillatory terms. They will not contribute
unless, as $q \rightarrow 0$, the phases are not zero or 
an integral multiple of $\pi$. In three dimensions, this will not 
occur unless there is a zero-energy resonance or a half-bound
state. In one dimension, this may occur even in the two-body
problem and a repulsive force\cite{all}, where we have 
found that $\delta(0) = -\pi/2$, for a delta function interaction.

\newpage
\begin{center}
\section*{Perspective}
\end{center}
{\bf Our phase shift formula makes sense to us.} 
\esp

In previous work, under the constraints of positive potentials
and Boltzmann statistics,
we were able to show that, in a semi-classical approximation,
the phase shift sums in our quantum mechanical formulation
lead to the correct classical expressions for the 3-body cluster.  
Each sum was shown to be 
associated with a classical expression for an integrand, 
involving the potential terms, which integrated over
the position variables, diverges as the volume increases to infinity.
When all of the terms of the cluster are taken together, the resulting
integral is finite and volume independent.
\esp

Terms in $\rho_{max}$ play no role in this, and in our previous work
using McGuire's model with repulsive potentials, as well as in the 
present work with attractive forces, we see that the $\rho_{max}$ 
terms cancel out. Unfortunately we cannot demonstrate this in general.
Since these terms are not tied to potentials, we feel that they are 
artifacts, of no fundamental importance.
\esp

{\bf Another aspect is of interest.} 
\esp

In an important paper, Mazo\cite{maz} noted that using the asymptotic
expression [here, argument much larger than the order] for the 
wave functions used to calculate the partition function of one
particle in a box (a sphere), leads to an incorrect result. 
`The important angular momenta $\ell$ are very large and 
increase proportionately with the size of the sphere.'
An improved calculation yields the correct result. 
\esp

Mazo determined, however, that the phase shift formula for the 
2-body cluster (virial) problem was correct, because the sum involved  
only a few angular momenta. The issue merits further discussion.
\esp

Both in the 2-body expressions and in ours, there is an integration 
over the energy (or the wave number). The effective range of the energy 
is limited by a Boltzmann factor, such that, for a given temperature 
and accuracy, there is an upper limit to the energy to be considered. 
\esp

For the 2-body system, for a finite range of the interaction, say `a', 
there is then a semi-classical argument, which can be extended appropriately
in the full quantum mechanics, which indicates that the `higher' angular 
momenta $\ell > k_{max} a$ are not involved in the
scattering (or in the full
quantum mechanics, for sufficiently large $\ell$, only in an exponentially
decreasing fashion). 

\esp
Now, in our case, and that is the intuitive argument presented 
in our first paper\cite{lam}, we presented an analogous argument 
for the 3-body cluster, which also has - in position space - a
finite extent (which is why it was devised).
I.e., the cluster has been constructed so that it only contributes 
when particles are involved in a truly three-body event.

\esp
In the 3-body problem, when describing the collision, 
the Global Angular Momentum ($K$ being the order) measures how 
closely three bodies simultaneously approach
each other\cite{smith}. A semi-classical point of view would then 
suggest that only a limited (or convergent) number of the higher K angular
functions would be required to correctly describe the properties
of the cluster, at a given energy.

\esp
Of course, our adiabatic basis is quite superior to the original 
hyperspherical basis used in our first paper, but asymptotically,
for 3-3 scattering, it reduces to hyperspherical harmonics. 
We think that the argument that we have presented is a dependable guide.

\esp
In our work with McGuire's model, we had the advantage of being
able to obtain analytical results. In our previous paper -
and in this one, in which we also use the same result - we evaluate
the difference of two sums, each over an infinite number of phase shifts,
but for which the difference converges. To obtain this difference,
we had to use a procedure due to Abel to help with the summations. 
\esp

For the third cluster in 2-dimension\cite{laz,tony}, analytical and numerical 
data considerations led to the vanishing of the leading part of the first 
Born contribution to the authors' equivalent to our $G(k)$, Eq.(24). 
This led to the correct threshold behaviour of the $G(k)$ and the 
definite demonstration of convergence for small values of $k$.
\footnote{SYL - as coauthor of this paper with Jei Zhen - deplores the 
rash extrapolation of the $G(k)$ in the figure, for which only he is
responsible. The remainder of the paper is excellent, solid and correct.} 
\esp

As we can see, there are sensitive issues and a combination of analytical
and numerical results is really required, if one is to take advantage
of the formalism that we have proposed and now extended.
\esp
 
Finally, we have always emphasized that our hyperspherical-adiabatic 
method could be extended to more particles (and dimensions).
Higher clusters, however, would mean more phase shift sums and more
subtractions!  
\esp

\newpage
\begin{center}
\section*{Conclusions}
\end{center}
The effort that led to this paper, is the last in a long sequence of 
efforts, starting with an early formulation of the third cluster
(or virial) in terms of hyperspherical harmonics, all in an attempt
to generalize the formula of Ulhenbeck and Beth, and of Gropper,
to the higher clusters and virials.
\esp

It led to formulations in the continuum, out of the box;  
to the consideration of the second virial with anisotropic interactions
\footnote{SYL found out - to his dismay - that online the reference
is associated with Yves instead of Larsen, and of course should 
also be attributed to his co-author Poll};
to the usefulness of an adiabatic approximation in using our
hyperspherical formalism, and then reformulations
in terms of an adiabatic basis. It led, importantly, within constraints,
to devising a WKB + adiabatic approximation,  a semi-classical approach, 
so as to obtain the classical expressions from the eigenphase shift 
formalism, as it existed then.
\esp

It led to work in two dimensions and to our obtaining (together with our
Russian friends) a wealth of analytical results 
(adiabatic basis, eigenpotentials, eigenphase shifts, 
W-matrix, S-matrix), for one-dimensional delta function models. 
\esp

Finally, now, in the present paper, we present our most elegant,
our most general result, our full generalization of the famous
Ulhenbeck and Beth formula. We have gone as far as we could.
In the previous section, Perspective, we have tried to draw attention
to sensitive aspects, and perhaps limitations, of the phase shift approach.
\esp

To complement the work, we have calculated explicitly, in an 
adiabatic approximation, the $b_3$ cluster for the attractive
version of McGuire's model. To our knowledge, this is also new.

\newpage
\begin{center}
{\bf Acknowlegments}
\end{center}
\vspace{.2cm}
Alejandro Amaya thanks the partial support, in the early stages of this work,
of the DGAPA, program PAPIIT-IN109511,
and Sigurd Larsen thanks the always welcoming 
hospitality of the Institutes, the ICF of Mexico and the IPN Orsay.

\esp
In this, the last of the long sequence of papers that have lead to 
our results, SYL wishes to thank his many collaborators over the years, 
and especially his present coauthors, who in friendship and intellectual 
support and contribution have been essential to the communal effort.
These papers would not have been possible without their help and 
contributions. The value of their friendship has been inestimable.

\newpage

\section*{Appendix A}
{\bf The eigenphase-shift eigenfunctions}  \\
\esp

In this Appendix we show, in a more detailed fashion than shown in
(\cite{lap}),
that we can choose solutions, for finite sets of coupled
equations from Eq. (\ref{eigen2}), such that
a unique eigenphase shift characterizes the asymptotic behaviour
of each of these solutions. We simplify the discussion,
in a manner appropriate to our section `Without Bound States'.
We append a `coda' to generalize our discussion to include 
bound states, and possible excited states in the asymptotic
states. The point is then, that for the following discussion,
the eigenpotentials go to zero, as $\rho$ approaches infinity. 

\esp
\esp
\noindent
{\bf Without Bound States and Excited States} 
\esp
\esp

Changing notation, so as to ultimately connect with asymptotic 
solutions $i$ of Eq. (\ref{eigen2}) for $\rho$
large, with components labeled by the index
$\ell$, we note that ALL the solutions of (\ref{eigen2}) can, 
for finite sets
(however large), and for sufficiently large values of $\rho$, be written
in the form
\begin{equation}
{\chi}_{\ell}^{\lambda}\left(k,\rho\right)\rightarrow\left(k\rho\right)^{1/2}
\left[A^{\lambda}_{\ell}\left(k\right)\, J_{K_{\ell}+2}\left(k\rho\right)+
B^{\lambda}_{\ell}\left(k\right)\, N_{K_{\ell}+2}\left(k\rho\right)\right],
\label{eq:a1}
\end{equation}
where $K_{\ell}$ is one of the quantum labels included in the index $\ell$,
and $\lambda$ denotes the solution.
The linear combinations,
\begin{equation}
\begin{array}{cl}
\zeta_{\ell}^{\mu}\left(k,\rho\right) & 
={\displaystyle {\displaystyle {\displaystyle 
{\displaystyle \sum_{\lambda}}}}\chi_{\ell}^{\lambda}\left(k,\rho\right)
\left[A^{-1}\right]^{\mu}_{\lambda}\left(k\right)}\\
\esp
 & \sim\left(k\rho\right)^{1/2}\left[\delta^{\mu}_{\ell}\, 
J_{K_{\ell}+2}\left(k\rho\right)+W_{\ell}^{\mu}\left(k\right)\, 
N_{K_{\ell}+2}\left(k\rho\right)\right],
\end{array}\label{eq:a2}
\end{equation}
are of particular interest, because the matrix $W$, with elements
defined by
\begin{equation}
W_{\ell}^{\mu}\left(k\right)={\displaystyle 
{\displaystyle {\displaystyle \sum_{\lambda}}B^{\lambda}_{\ell}\left(k\right)}}
\left[A^{-1}\right]^{\mu}_{\lambda}\left(k\right),\label{eq:a3}
\end{equation}
is symmetric (as it is shown in Appendix B). In our case, the $W$
matrix is also real, so it can be diagonalized by a real orthogonal 
matrix ${\mathcal{C}}$, leading to a unique eigenphase shift,
for each solution,
associated with all components of each of the new solutions. This important
property is demonstrated by multiplying the functions
defined in Eq. (\ref{eq:a2}) by the matrix elements of ${\mathcal{C}}$.
Then by using the definition of the orthogonal matrix
\[
{\displaystyle \sum_{i'}}{\mathcal{C}}^{i'}_{\ell}\left(k\right)
\left[{\mathcal{C^T}}\right]^{\mu'}_{i'}\left(k\right)=\delta^{\mu'}_{\ell}
\]
and defining the eigenphase shift in terms of the $W$ eigenvalues
as
\[
{\displaystyle \sum_{\mu'}}\left[{\mathcal{C}}^{T}\right]^{\mu'}_{i'}
\left(k\right)\left[\sum_{\mu}{\displaystyle 
W_{\mu'}^{\mu}\left(k\right)} \,
{\mathcal{C}}^{i}_{\mu}\left(k\right)\right]=-\delta_{i'}^{i} \, 
\tan\left(\delta^{i}(k)\right),
\]
we obtain the desired solution, with a unique eigenphase shift shared by
each of its components,
\[
\begin{array}{cl}
\tilde{\phi}_{\ell}^{i}\left(k,\rho\right) & 
={\displaystyle \sum_{\mu}}{\displaystyle 
\zeta_{\ell}^{\mu}\left(k,\rho\right)} \,  \,
{\mathcal{C}}^{i}_{\mu}\left(k\right)\\
\\
 & \sim\left(k\rho\right)^{1/2}{\mathcal{C}}^{i}_{\ell}\left(k\right)\,
\left[J_{K_{\ell}+2}\left(k\rho\right)-\tan\left(\delta^{i}(k)\right)\, 
N_{K_{\ell}+2}\left(k\rho\right)\right].
\end{array}
\]

\esp
\esp
\noindent
{\bf With Bound States, or/and Excited States}
\\

\esp
\noindent
Essentially, we have the same type of asymptotic formulae:
\esp
\noindent
\[
\begin{array}{cl}
\tilde{\phi}_{\ell}^{i}\left(q_{\alpha},\rho\right) &
={\displaystyle \sum_{\mu}}{\displaystyle
\zeta_{\ell}^{\mu}\left(q_{\alpha},\rho\right)} \,  \,
{\mathcal{C}}^{i}_{\mu}\left(q_{\alpha}\right) \\
\\
 & \sim\left(q_{\alpha}\rho\right)^{1/2}{\mathcal{C}}^{i}_{\ell}
\left(q_{\alpha}\right)\,
\left[J_{K_{\ell}+2}\left(q_{\beta}\rho\right)-\tan\left(\delta^{i}(q_
{\alpha})\right) \,
N_{K_{\ell}+2}\left(q_{\beta}\rho\right)\right].\label{eq:a4}
\end{array}
\]
but the $q_{\beta}$ depends on the kinetic energy, which,
asymptotically, we find in the fragment channels, 
and the $q_{\alpha}$ which depends on what the integration 
variable is over the energy, such that each amplitude has a 
delta function normalization. We refer to our ref(\cite{lap}).

\newpage
\section*{Appendix B}
In this appendix we show that the matrix $W$ (\ref{eq:a3}) is symmetric,
using the same approach followed in reference \cite{lap}.\esk

\esp
\esp
\noindent
{\bf Without Bound States and Excited States}
\esp
\esp

Let us consider a solution $\zeta^{\mu}$ of Eq. (\ref{eigen2}) (see Eq. (\ref{eq:a2})). Then from the relation
\begin{equation}
{\displaystyle \sum_{\ell}}\left[\zeta_{\ell}^{\mu}\left(k,\rho\right){\cal O}
\left(k',\rho\right)\zeta_{\ell}^{\mu'}\left(k',\rho\right)-\zeta_
{\ell}^{\mu'}\left(k',\rho\right){\cal O}\left(k,\rho\right)\zeta_{\ell}^{\mu}
\left(k,\rho\right)\right]=0,\label{eq:b1}
\end{equation}
where
\begin{equation}
\begin{array}{cl}
{\cal O}\left(k,\rho\right)\zeta_{\ell}^{\mu}\left(k,\rho\right) & 
=\left(\frac{d^{2}}{d\rho^{2}}-\Lambda_{\ell}\left(\rho\right)+
k^{2}\right)\zeta_{\ell}^{\mu}\left(k,\rho\right)\\
 & +2{\displaystyle \sum_{\ell'}}C_{\ell}^{\ell'}\left(\rho\right)
\frac{d}{d\rho}\zeta_{\ell'}^{\mu}\left(k,\rho\right)\\
 & +{\displaystyle 
\sum_{\ell'}}D_{\ell}^{\ell'}\left(\rho\right)\zeta_{\ell'}^{\mu}
\left(k,\rho\right),
\end{array}\label{eq:b2}
\end{equation}
we obtain the identity
\begin{equation}
\begin{array}{l}
{\displaystyle \sum}\frac{d}{d\rho}\left[\zeta_{\ell}^{\mu}
\left(k,\rho\right)
\frac{d}{d\rho}\zeta_{\ell}^{\mu'}\left(k',\rho\right)-
\zeta_{\ell}^{\mu'}
\left(k',\rho\right)\frac{d}{d\rho}\zeta_{\ell}^{\mu}\left(k,\rho\right)
\right]\\
\ell\\
=\left\{\left(k'{}^{2}-k^{2}\right){\displaystyle \sum_{\ell}}\zeta_
{\ell}^{\mu'}\left(k',\rho\right)\zeta_{\ell}^{\mu}\left(k,\rho\right)\right.\\
\\
\left.
+2{\displaystyle \sum_{\ell,\ell'}}
\frac{d} {d\rho}\left[\zeta_{\ell}^{\mu}\left(k,\rho\right) \,
C_{\ell}^{\ell'} \left(\rho\right) \,
\zeta_{\ell'}^{\mu'}
\left(k',\rho\right)\right]\right\}.
\end{array}\label{eq:b3}
\end{equation}
where, for the last term, we used Eq.(\ref{anti}).
Integrating over $\rho$ when $k=k'$ leads us to the following equation, 
\begin{equation}
\begin{array}{l}
{\displaystyle \sum_{\ell}}\left[\zeta_{\ell}^{\mu}\left(k,\rho\right)
\frac{d}{d\rho}\zeta_{\ell}^{\mu'}\left(k,\rho\right)-
\zeta_{\ell}^{\mu'}\left(k,\rho\right)\frac{d}{d\rho}\zeta_{\ell}^{\mu}
\left(k,\rho\right)\right]_{\rho=\rho_{max}}=0.
\end{array}\label{eq:b4}
\end{equation}
We used the fact that $\zeta_{\ell}^{\mu}$ goes to zero
as $\rho$ itself goes to zero, and that $C_{\ell}^{\ell'}$ decreases
fast enough for $\rho$ large. We can then substitute, in the above
expression, the asymptotic
form of the solutions, Eq. (\ref{eq:a2})
in appendix A, valid for large $\rho$. We obtain
\begin{equation}
\begin{array}{cl}
0= & k\rho\left(W_{\mu}^{\mu'}\left(k\right)-W_{\mu'}^{\mu}\left(k\right)\right)\,\left(J_{K_{\ell}+2}\left(k\rho\right)\,\frac{d}{d\rho}\left[\left(k\rho\right)^{1/2}N_{K_{\ell}+2}\left(k\rho\right)\right]\right.\\
\\
 & \left.-N_{K_{\ell}+2}\left(k\rho\right)\frac{d}{d\rho}\left[\left(k\rho\right)^{1/2}J_{K_{\ell}+2}\left(k\rho\right)\right]\right)
\end{array}\label{eq:b5}
\end{equation}
The evaluation of the Wronskian for the Bessel's functions \cite{abs},
leads to the equality
\begin{equation}
W_{\mu}^{\mu'}\left(k\right)=W_{\mu'}^{\mu}\left(k\right),\label{eq:b6}
\end{equation}
which proves that the matrix $W$ is symmetric.

\esp
\esp
\noindent
{\bf With Bound States, or/and Excited States}
\\

\esp
\noindent
The appropriate asymptotic formulae would change the Eq.(\ref{eq:b5}) 
into the following equation:
\begin{equation}
\begin{array}{cl}
0= & q_{\alpha}\rho\left(W_{\mu}^{\mu'}\left(q_{\alpha}\right)-W_{\mu'}^{\mu}\left(q_{\alpha}\right)\right)\,\left(J_{K_{\ell}+2}\left(q_{\beta}\rho\right)\,\frac{d}{d\rho}\left[\left(q_{\alpha}\rho\right)^{1/2}N_{K_{\ell}+2}\left(q_{\beta}\rho\right)\right]\right.\\
\\
 & \left.-N_{K_{\ell}+2}\left(q_{\beta}\rho\right)\frac{d}{d\rho}\left[\left(q_{\alpha}\rho\right)^{1/2}J_{K_{\ell}+2}\left(q_{\beta}\rho\right)\right]\right)
\end{array}\label{eq:b7}
\end{equation}
where $q_{\alpha}$ and $q_{\beta}$ have the same meaning as in Appendix A. 
From the above relation we can prove the same symmetric property, 
Eq. (\ref{eq:b6}), for the matrix W.

\newpage
\section*{Appendix C}
\noindent
In this Appendix we develop formulae associated to the four
lowest eigensolutions of Eq.(\ref{eigen1}), corresponding
to $K=0,3$ (Cosine basis) and $K=0,1$ (Sine basis), for the system
of three particles on a line interacting through delta function potentials.
A few of them appeared in our previous works. See \cite{alp}, Eqs. (26,33,34) 
and also \cite{cgk}, Eqs. (A1).

\subsection*{Cosine basis}

\noindent For $K=0$ the adiabatic function basis reads :
\begin{equation}
B_{0}(\rho,\theta)=\sqrt{N}\ \ \cos(q_{0}\theta),\label{C1}
\end{equation}
where
\begin{equation}
\begin{array}{rllll}
\theta\in[-\pi/6,\pi/6]\qquad & \mathrm{and}\qquad q_{0}\tan\left(q_{0}\frac{\pi}{6}\right) & =-\frac{\pi\rho}{6}\: & {\rm for} & {\rm H_3},\\
\\
\theta\in[-\pi/2,\pi/2]\qquad & \mathrm{and}\qquad 
q_{0}\tan\left(q_{0}\frac{\pi}{2}\right) & =-\frac{\pi\rho}{6}\: 
& {\rm for} & {\rm (H_2+T_1)},
\end{array}\label{C2}
\end{equation}
and the normalization factors may be written as
\begin{equation}
\begin{array}{rlll}
N & =\int_{-\pi/6}^{\pi/6}\cos^{2}(q_{0}\theta)\ {\rm d}\theta & =\frac{\pi}{6}\left(1-\frac{\rho}{q_{0}^{2}+\pi^{2}\rho^{2}/36}\right) & {\rm for\ H_3},\\
\\
N & =\int_{-\pi/2}^{\pi/2}\cos^{2}(q_{0}\theta)\ {\rm d}\theta & 
=\frac{\pi}{2}\left(1-\frac{\rho/3}{q_{0}^{2}+\pi^{2}\rho^{2}/36}\right) & 
{\rm for (\ H_2+T_1)}\ .
\end{array}\label{C3}
\end{equation}
We observe that the equations in (\ref{C2}) imply that
\begin{equation}
(\forall\rho)\qquad\quad\ q_{0}^{(H_2+T_1)}(\rho)=q_{0}^{(H_3)}(3\rho)/3\ .\label{C4}
\end{equation}
\medskip{}
The introduction of the above relation in the definitions of  $\Delta$ and  
$\Lambda$,
\begin{equation}
\Lambda_{0}(\rho)=\Delta_0-\frac{1/4}{\rho^{2}}; \quad \Delta_0=
\frac{q_{0}^{2}(\rho)}{\rho^{2}}\label{C5}
\end{equation}
\medskip{}
\noindent
and in the definition of $D$ (see Eq. (\ref{Dmatrix})) leads us to 
write the relations
beetwen variables in the cases, $H_3\,\mathrm{and\,}(H_2+T_1)$ as:
\begin{equation}
\begin{array}{rlll}
(\forall\rho) &\qquad\quad\Delta_{0}^{(H_2+T_1)}(\rho) &=&\Delta_{0}^{(H_3)}(3\rho)
\label{C60}\\
\\
(\forall\rho) &\qquad\quad\Lambda_{0}^{(H_2+T_1)}(\rho) &=&\Lambda_{0}^{(H_3)}(3\rho)-\frac{2}{\left(3\rho\right)^{2}}\label{C6}\\
\\
(\forall\rho) &\qquad\quad\ D_{0,0}^{(H_2+T_1)}(\rho) &=&9D_{0,0}^{(H_3)}(3\rho)\ .
\end{array}\label{prop3}
\end{equation}
\noindent Next, for small $\rho$ we collect the expansions in powers
of $\rho$ for $q_0$, $\Lambda_0$ and $D_{0,0}$ in the case of $H_3$:\medskip{}
\begin{eqnarray}
q_{0}(\rho) & = & i\sqrt{\rho}\left(1+\frac{\pi^{2}}{216}\ \rho+\frac{11\pi^{4}}{466560}\ \rho^{2}+\frac{17\pi^{6}}{235146240}\ \rho^{3}-\frac{281\pi^{8}}{1015831756800}\ \rho^{4}\right.\nonumber \\
\nonumber \\
 &  & \left.-\frac{44029\pi^{10}}{7240848762470400}\ \rho^{5}+\dots,\right)\label{C8a}\\
\nonumber \\
\Lambda_{0}(\rho) & = & -\frac{1}{4\rho^{2}}-\frac{1}{\rho}-\frac{\pi^{2}}{108}-\frac{\pi^{4}}{14580}\ \rho-\frac{\pi^{6}}{2755620}\ \rho^{2}-\frac{\pi^{8}}{1488034800}\ \rho^{3}\nonumber \\
\nonumber \\
 &  & +\frac{\pi^{10}}{88389267120}\ \rho^{4}+\ldots,\label{C8b}\\
\nonumber \\
D_{0,0}(\rho) & = & -\frac{\pi^{4}}{58320}-\frac{\pi^{6}}{2204496}\ \rho-\frac{31\ \pi^{8}}{5952139200}\ \rho^{2}-\frac{\pi^{10}}{117852356160}\ \rho^{3}\nonumber \\
\nonumber \\
 &  & \left.+\frac{1151\ \pi^{12}}{1608684661584000}\ \rho^{4}+\ldots\right)\label{C8c}
\end{eqnarray}
\medskip{}
 The corresponding expansions for large $\rho$ would be:
\begin{eqnarray}
q_{0}\left(\rho\right) & =&\frac{6i}{\pi}\ (\alpha+2\alpha\exp(-2\alpha)),\qquad\quad\alpha=\frac{\pi^{2}\rho}{36}; \nonumber\\ 
& & \nonumber\\
\Lambda_{0}\left(\rho\right) & =&-\frac{\pi^{2}}{36}-\frac{1}{4\ \rho^{2}}-\frac{\pi^{2}}{9}\exp(-\pi^{2}\rho/18)+\dots,
\label{C9}\\
& &  \nonumber\\
D_{0,0}\left(\rho\right) & =& -\frac{1}{4\ \rho^{2}}-\pi^{2}\left(\frac{\pi^{4}\ \rho}{17496}-\frac{\ \pi^{2}}{324}-\frac{1}{18\ \rho}\right)\exp(-\pi^{2}\rho/18)+\dots \nonumber
\end{eqnarray}
and the analogous expressions for both, small $\rho$ and $\rho$ large,
in the case of $(H_2+T_1)$ can be obtained from Eqs. (\ref{C4}, \ref{C6} and
\ref{prop3} ).

\noindent 
Note that the diagonal part of the asymptotic adiabatic
interactions, $\Lambda_{0}\left(\rho\right)-D_{0,0}\left(\rho\right)$,
approaches exponentially the two-body bound energy $-\pi^{2}/36$.
(See Fig. 4).

\subsection*{Sine basis}

\noindent The adiabatic basis reads :
\begin{equation}
B_{K}(\rho,\theta)=\sqrt{N}\ \sin(q_{K}\theta)\label{C10}
\end{equation}
where
\begin{equation}
\begin{array}{rllll}
\theta\in[-\pi/6,\pi/6]\; & \mathrm{and}\ q_{3}\cot\left(q_{3}\frac{\pi}{6}\right) & =-\frac{\pi\rho}{6} & {\rm for} & {\rm H_3},\\
\\
\theta\in[-\pi/2,\pi/2] & \;\mathrm{and\;}q_{1}\cot\left(q_{1}
\frac{\pi}{2}\right) & =-\frac{\pi\rho}{6}, & {\rm for} & {\rm (H_2+T_1)},
\end{array}\label{C11}
\end{equation}
and the normalization factors may be written as
\begin{equation}
\begin{array}{rlll}
N & =\int_{-\pi/6}^{\pi/6}\cos^{2}(q_{3}\theta)\ {\rm d}\theta & =\frac{\pi}{6}\left(1-\frac{\rho}{q_{3}^{2}+\pi^{2}\rho^{2}/36}\right) & {\rm for\ H_3},\\
\\
N & =\int_{-\pi/2}^{\pi/2}\cos^{2}(q_{1}\theta)\ {\rm d}\theta & 
=\frac{\pi}{2}\left(1-\frac{\rho/3}{q_{1}^{2}+\pi^{2}\rho^{2}/36}\right) & 
{\rm for\ (H_2+T_1)}\ .
\end{array}\label{C12}
\end{equation}

\noindent The equations in (\ref{C11}) imply that
\begin{equation}
(\forall\rho)\qquad\quad q_{1}^{(H_2+T_1)}(\rho)=q_{3}^{(H_3)}(3\rho)/3\ .\label{C13}
\end{equation}
Hence, taking into account that
\begin{equation}
\Delta_{K}=\frac{q_{K}^{2}(\rho)-K^{2}}{\rho^{2}};\quad K>0,\label{C140}
\end{equation}
\begin{equation}
\Lambda_{K}=\frac{q_{K}^{2}(\rho)-1/4}{\rho^{2}};\quad K>0,\label{C14}
\end{equation}
and the definition of the matrix $D$ (Eq.(\ref{Dmatrix})), we can write:

\begin{eqnarray}
(\forall\rho)\ \qquad\quad\Delta_{1}^{(H_2+T_1)}(\rho) & =&
\Delta_{3}^{(H_3)}(3\rho) \ , \label{C150}\\
  & & \nonumber\\
(\forall\rho)\ \qquad\quad\Lambda_{1}^{(H_2+T_1)}(\rho) & =&
\Lambda_{3}^{(H_3)}(3\rho)-\frac{2}{\left(3\rho\right)^{2}} \ , \label{C15}\\
& &  \nonumber\\
(\forall\rho)\qquad\quad\ D_{1}^{(H_2+T_1)}(\rho) & =& 9D_{3}^{(H_3)}(3\rho)\ .
\label{C151}\end{eqnarray}

\noindent For small $\rho$  the expansions in powers of $\rho$ for
$q_{3}$, $\Lambda_{3}$ and $D_{3,3}$ in the case of $H_3$ are:
\noindent
\begin{eqnarray}
q_{3} & = & 3-\frac{1}{3}\ \rho-\frac{1}{27}\ \rho^{2}+\frac{(\pi^{2}-24)}{2916}\ \rho^{3}+\frac{(\pi^{2}-15)}{6561}\ \rho^{4}\nonumber \\
\nonumber \\
 &  & -\frac{(1120-100\pi^{2}+\pi^{4})}{1574640}\ \rho^{5} -\frac{(10080-1120\ \pi^{2}+23\ \pi^{4})}{42515280}\ \rho^{6}+\dots,\label{c16}\\
\nonumber \\
\Lambda_{3} & = & \frac{35}{4\ \rho^{2}}-\frac{2}{\rho}-\frac{1}{9}+\frac{\pi^{2}-12}{486}\ \rho+\frac{\pi^{2}-10}{1458}\ \rho^{2}-\frac{1680-200\ \pi^{2}+3\ \pi^{4}}{787320}\ \rho^{3}\nonumber \\
\nonumber \\
 &  & -\frac{6048-840\ \pi^{2}+23\ \pi^{4}}{8503056}\ \rho^{4}+\dots,\label{c17}\\
\nonumber \\
D_{3,3} & = & -\frac{\pi^{2}-9}{972}-\frac{(4\ \pi^{2}-39)}{4374}\ \rho+\frac{(978-109\ \pi^{2}+\pi^{4})\,\rho^{2}}{157464}\ \rho^{2}\nonumber \\
\nonumber \\
 &  & +\frac{(16092-2025\ \pi^{2}+40\ \pi^{4})}{4251528}\ \rho^{3}\nonumber \\
\nonumber \\
 &  & -\frac{(-656352+93624\ \pi^{2}-2837\ \pi^{4}+9\ \pi^{6})}{306110016}\ \rho^{4}+\dots\label{c18}
\end{eqnarray}

\noindent The corresponding expansions for large $\rho$ would be
\begin{eqnarray}
q_{3} & = & \frac{6i}{\pi}\ (\alpha-2\alpha\exp(-2\alpha))\qquad\quad\alpha=\frac{\pi^{2}\rho}{36},\\
\nonumber\\
\Lambda_{3} & = & -\frac{\pi^{2}}{36}-\frac{1}{4\ \rho^{2}}+\frac{\pi^{2}}{9}\exp(-\pi^{2}\rho/18)+\dots\\
\nonumber\\
D_{3,3} & =- & \frac{1}{4\ \rho^{2}}+\pi^{2}\left(\frac{\pi^{4}\ \rho}{17496}-\frac{\pi^{2}}{324}-\frac{1}{18\ \rho}\right)\exp(-\pi^{2}\rho/18)+\dots,
\end{eqnarray}
and the analogous expressions for both, small $\rho$ and large $\rho$,
in the case of $(H_2+T_1)$ can be obtained from 
Eqs. (\ref{C13}, \ref{C5} and \ref{C151}).

\esp
Again, as in the `Cosine' case, the diagonal part of the asymptotic 
adiabatic interactions, 
$\Lambda_{K}\left(\rho\right)-D_{K,K}\left(\rho\right)$, 
approaches the two-body bound energy $-\pi^{2}/36$ exponentially, as
we can see that the $1/(4 \rho^2)$ cancels. 
(See also Fig. 4).

\esp
This is important. This implies that, in ALL four cases,  
the analysis of the asymptotic form of the solution of the relevant 
Schr\"odinger equation:
\beq 
\left( \frac{d^2}{d\rho^2} -\Lambda_K + D_{K,K}+q^2-\frac{\pi^2}{36} 
\right) \phi_K(q,\rho)=0
\eeq
will involve Bessel and Neumann functions of order $1/2$  \dots
leading to  simple asymptotic formulations of the form
$\sin(q \rho + \delta)$ 
for all of these cases. We have used this in obtaining our phase
shifts and also in discussing the contribution of the oscillatory
terms, Eq.(\ref{osc}).

\newpage
\section*{Appendix D}
\noindent
In their (2005) article\cite{mes}, Mehta and Shepard write that 
their phase shifts "differ in a critical way" from those presented 
in our work\cite{alp}. Further they assert that "the definition of our
S-matrix is consistent with the threshold behavior 
of the effective range expansion and with the statement of Levinson's theorem
in one dimension". We wish to respond.
\esp

We first would like to exhibit our adiabatic $H_2 + T_1$ phase shift, as 
obtained from our eigenpotential + diagonal coupling element, and 
values from our understanding of the exact phase shift, based on our
evaluation of the phase, \`{a} la McGuire\cite{mcg,alp}.

\begin{figure}[h]
\input{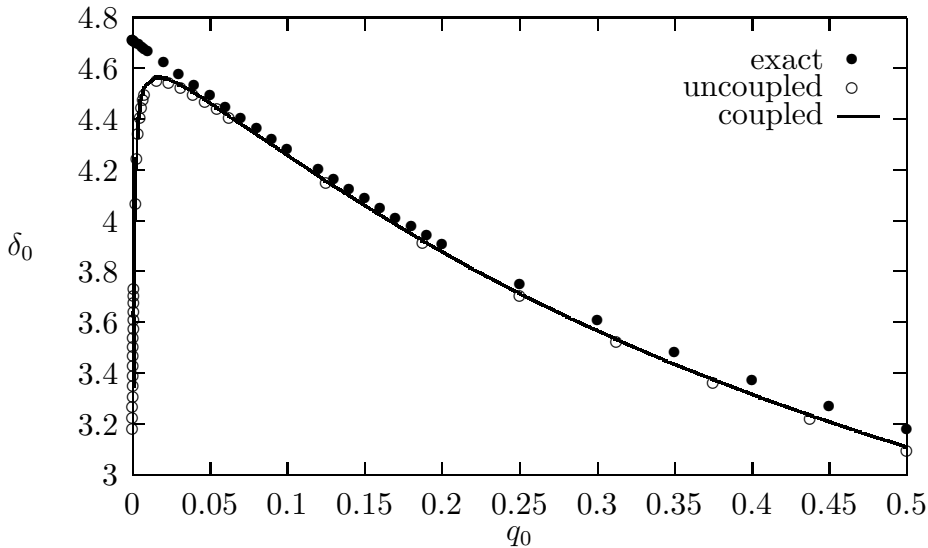}
\caption{The exact and the adiabatic $H_2+T_1$ phase shift}
\end{figure}
We remark that our numerical results, which we obtained both by solving
the Schr\"{o}dinger and the Riccati equations, will when using the phase 
equation\cite{cal} {\em automatically} incorporate the factors
of $\pi$'s, associated with bound states. The phase formalism can be used
to prove Levinson's theorem, including the zero energy resonances. 
See the book by Calogero, cited in the last reference, Chapter 22. 
We see that our numerical results (for one and 2 coupled equations) 
indicate an enormous scattering
length  for our eigenpotential (+ diagonal coupling term). 
We note that we are calculating for a potential which is
an upperbound, but close to, the potential which would yield the exact answer.
This implies the
correctness of describing a phase shift by an expression which
yields $3 \pi/2$ as the value at zero energy.
\esp

Mehta and Shepard state that our $3 \pi/2$ would be consistent with 
Levinson's theorem in three dimensions, but not in one. They quote results 
valid in one dimension, but for the two (one, since we factor the 
c.m. motion) particle problem, with a range of the distance from 
$ - \infty$ to $\infty$!
Our three particle problem is closer to the three dimension situation than
to that of one dimension. Our formalism, as that borrowed by M\&S, involves 
hyperspherical potentials and radial equations!
\esp
\newpage
\esp

To eliminate the (from their point of view) spurious additional factor
of $\pi/2$, they change the sign of the S-matrix (!), thereby writing
their basic wave function (mod a factor of $1/\sqrt{\rho}$), as 
$\cos(q\rho + \delta_{MS})$, instead of $\sin(q\rho + \delta)$ .
They also do not use the conventional effective range formula:
instead of using the cotangent in 
$q \cot(\delta) = 1/\alpha + r_0 q^2/2 ...$, they use the tangent,
in the similar formula, as seen in the fifth line below their Eq.(12).
If they were to use the conventional effective range formula,
they would find that their $r_0 = 16\sqrt{3}/\pi$.
\esp

We suspect a sign error in their
S-matrix, and therefore an error in their $\tan(\delta)$.
For $q = 0$, $\tan(\delta)$ should be infinite.
\esp

We note that their phase shift merely differs from ours by $\pi/2$.
I.e., $\delta_{MS} =  \delta - \pi/2$. Since Levinson's theorem 
involves the difference between the values of the phase shifts
at the origin and at infinity, a constant should not matter.
We would like to emphasize that their argument that the change in the
sign of the S-matrix is `due' to the fact that asymptotically we have a 
2-body bound state and a `free' particle does not stand up.
Any oscillating solution of the radial equation for this eigenpotential - 
which has a `plateau' at large distance - is associated to an adiabatic
function, which at large distances (and small angles) reduces to 
a bound state solution of the 2-body problem. The issue of the $\pi/2$ 
is irrelevant, and we, as well as they, are certainly aware of the 
resonance at the 2-body bound state energy!
\esp

In our work we need only the derivative of the 
phase shifts.
Since, however, we need the threshold behaviour of the phase 
illustrated in our Figure 6, and basing ourself on our fundamental
result of Eq.(60) in our `old' paper, we proceed as follows. 
We assert that 
our `old' expression, Eq.(61) is equal to {\em minus} 
the exact S-matrix, and for small values of $q$, 
we therefore expanded  (61), multiplied it by $e^{i 3 \pi}$,
and took $1/2i$ times the logarithm.
We obtained:
\beq
\delta \sim \frac{3\pi}{2} -  \frac{8\sqrt{3}}{\pi}\, q \ldots
\eeq
and additional numerical results. \\
A nicer formula in terms of real variables was obtained by our Russian 
colleagues\cite{vlp} in their Eq.(54):
\beq
\delta_{exact} = \frac{3\pi}{2} - \arctan{\frac{8\sqrt{3}q/\pi}
{1 - 36 q^2/\pi^2}}
\eeq
Further, they, in two papers\cite{vlp, gcp} developed  an
`Effective Adiabatic Approach' - based on a `Canonical Asymptotic
Transformation' - which yields numerical values which are `spot-on'
the exact results.
\esp

Finally, since this is our opportunity, we would like to signal a 
misprint in our old $H_2+T_1$ paper. Eq.(57), should read:
\beq
   B_{0}(\rho,\vartheta) \sim \frac{\sqrt{\pi \rho}}{6}  e^{-
\frac{\pi} {6}\rho(\frac{\pi}{6} - |\vartheta - \frac{m\pi}{3}|)}.
\eeq
The variable $\rho$ was missing!

\end{document}